\documentclass[structabstract]{aa}  
\pdfoutput=1
\usepackage{txfonts}
\usepackage{longtable,lscape}
\usepackage{graphicx}
\usepackage{natbib}

\newcommand{\niia}{[N\,\textsc{ii}]6584}

\newcommand{\sii}{[S\,\textsc{ii}]6717,31}

 \newcommand{\ha}{H$\alpha$}
\newcommand{\msun}{M$_{\odot}$} 
\newcommand{\un}{$^{-1}$} \newcommand{\car}{$^{-2}$}
\newcommand{\fig}[1]{Fig.~\ref{#1}}
\begin{document}
\title{MASSIV: Mass Assembly Survey with SINFONI in VVDS\thanks{This work is based  
    on observations collected at the European Southern Observatory
    (ESO) Very Large Telescope, Paranal, Chile, as part of the Programs 
    179.A-0823, 78.A-0177, and 75.A-0318. This work also benefits from data products 
    produced at TERAPIX and the Canadian Astronomy Data Centre as part of the 
    Canada-France-HawaiiTelescope Legacy Survey, a collaborative project of NRC 
    and CNRS. 
}}
    \subtitle{III. Evidence for positive metallicity gradients in $z\sim1.2$ star-forming galaxies}

\author{
J. Queyrel \inst{1,2}
\and T. Contini \inst{1,2}
\and M. Kissler-Patig \inst{3}
\and B. Epinat \inst{1,2}
\and P. Amram \inst{4}
\and B. Garilli \inst{5}
\and O. Le F\`evre \inst{4}
\and J. Moultaka \inst{1,2}
\and L. Paioro \inst{5}
\and L. Tasca \inst{4}
\and L. Tresse \inst{4}
\and D. Vergani \inst{6}
\and C. L\'opez-Sanjuan \inst{4}
\and E. Perez-Montero \inst{7}
         }

\offprints{T. Contini, \email{contini@ast.obs-mip.fr}}

\institute{
  Institut de Recherche en Astrophysique et Plan\'etologie (IRAP), CNRS, 14, avenue Edouard Belin, F-31400 
Toulouse, France \and
  IRAP, Universit\'e de Toulouse, UPS-OMP, Toulouse, France \and
  ESO, Karl-Schwarzschild-Str.2, D-85748 Garching b. M\"unchen,
  Germany \and
  Laboratoire d'Astrophysique de Marseille, Universit\'e de Provence,
  CNRS, 38 rue Fr\'ed\'eric Joliot-Curie, F-13388 Marseille Cedex 13,
  France \and
  IASF-INAF, Via Bassini 15, I-20133, Milano, Italy \and
  INAF-Osservatorio Astronomico di Bologna, via Ranzani 1, I-40127,
  Bologna, Italy \and
  Instituto de Astrof\'\i sica de Andaluc\'\i a - CSIC Apdo. 3004
  E-18080, Granada, Spain
}

   \date{Received ; accepted }

\abstract{
  A key open issue for galaxy evolution and formation models is the understanding 
  of the different mechanisms of galaxy assembly at various cosmic epochs. 
  Spatially-resolved spectroscopy of high-redshift galaxies has proven to be a powerful 
  observing technique to address this issue.}{
  We aim at deriving the global and spatially-resolved metal content
  in high-redshift galaxies. This allows to study, mainly \emph{via} the
  determination of radial abundance gradients in the disks, their star formation
  history and their interaction with the surrounding intergalactic medium.}{
  Using VLT/SINFONI integral-field spectroscopy of a first sample of $50$~galaxies at $z\sim1.2$ in the MASSIV survey, we are able to measure the \ha{} and \niia{}
  emission lines. Using the N2 ratio as a proxy for oxygen abundance in the interstellar medium, we measure the 
  metallicity of the sample galaxies. We  develop a tool to
  extract spectra in annular regions of these galaxies, leading to a
  spatially-resolved estimate of the oxygen abundance in each
  galaxy.}{
  We derive a metallicity gradient for 26 galaxies in our sample and
  discover a significant fraction of galaxies with a ``positive''
  gradient.  Using a simple chemical evolution model, we derive infall
  rates of pristine gas onto the disks. }{
  Seven galaxies display a positive gradient at a high confidence level.
  Four out of these are interacting and one is a chain
  galaxy. We suggest that interactions might be responsible for
  shallowing and even inverting the abundance gradient.
  We also identify two interesting correlations in our sample: a) galaxies
  with higher gas velocity dispersion have shallower/positive
  gradients; and b) metal-poor galaxies tend to show a positive
  gradient whereas metal-rich ones tend to show a negative
  one. This last observation can be explained by the infall of
  metal-poor gas into the center of the disks. We address the question
  of the origin of this infall under the influence of gas flows
  triggered by interactions and/or cold gas accretion. 
  }


\maketitle
\section{Introduction}

An key open issue for galaxy evolution and formation models is the understanding of the different mechanisms of galaxy assembly at various cosmic epochs. In this context, the gas-phase and stellar metallicities have proven to
be important parameters to constrain the star formation history of galaxies. 
 
A relation between galaxy mass and metallicity, first discovered by \citet{leq79} in irregular galaxies, exists for star-forming galaxies both in the nearby universe \citep{tremonti04, lam04}, and at high redshifts $z\sim0.7-3$ \citep{lam06, lam09, perez09, queyrel09, erb06, maiolino08, manucci09}.  This trend for more massive galaxies
to have a higher gas-phase metallicity can be explained by various related properties: gas inflows and/or outflows, strength of the gravitational potential, efficiency of the star formation depending on stellar mass, etc. Their relative importance is currently being explored in numerical simulations.  

Studying the abundance in H\,\textsc{ii} regions of nearby spiral galaxies has unveiled metallicity gradients in the local Universe \citep{pageled81, vilacostas92, considere00, pilyugin04}. In $z\sim0$ galaxies, the metallicity generally decreases from the center to the outskirts (up to $\Delta_r Z\sim-0.1$ dex kpc\un{}).  
The strength of these gradients seems to correlate with various parameters such as rotation velocity,
morphological type, luminosity, and the presence of a bar. Different physical processes have been proposed to explain these patterns, including radial gas-flows \citep{koeppen94}, self-regulating star formation \citep{phillipps91}, variable yields \citep{vilacostas92}, or continuous infall of gas onto the disk \citep{magrini07}. 

Merging events \--- which are believed to play a substantial role in galaxy evolution  \citep[eg.][]{deravel09, lopez11}  \--- seem to be a key physical process in shaping the metallicity gradients of interacting galaxies \citep{rupke10}. Recent observations have suggested that galaxies involved in merging events show lower nuclear metallicities due to the infall of pristine gas into the nucleus. Merging events could account also for outliers to the mass-metallicity relation \citep{kewley06, md08, peeples09, queyrel09, alonso10, montuori10, kewley10}. 

Metallicity gradients have been observed in the stellar populations of early-type galaxies as well. They appear to correlate with macroscopic properties such as, for example, stellar mass  \citep[eg.][]{spolaor09}.  
It has been demonstrated that a metallicity gradient in the stellar populations could survive a major merger event, although it will be weakened. These observations contrast the predictions of the simple monolithical collapse scenario \citep{white80, bekki99, koba04, dimatteo09}.

In nearby galaxies, metallicity gradients can be inferred from the observation of different emission-line ratios (in H\,\textsc{ii} regions), or from stellar absorption lines. In the distant universe ($z\gtrsim0.5$), low signal-to-noise ratio (hereafter SNR) and low spatial-resolution data make these measurements much more challenging.  
However, thanks to powerful integral-field spectrographs mounted on the largest telescopes, it is now possible to determine spatially-resolved physical parameters of high redshift galaxies.  

Many studies have taken advantage of this technique to study the gas dynamics of distant galaxies, from $z\sim 0.6$ \citep{flores06} up to $z\gtrsim 2$ \citep{forster09, law09, epinat09, gnerucci11}. Recently, the Mass Assembly Survey with SINFONI in VVDS \citep[MASSIV;][]{contini11} collected data of $z\sim1-2$ galaxies observed with the integral-field spectrograph SINFONI at the VLT \citep{sinfo03}.  Depending on the galaxy redshift, the bright \ha{} emission line is targeted in the $J$ or $H$ band. The nitrogen line \niia{} close to  \ha{} allows to estimate the gas-phase metallicity \emph{via} the N2 ratio ($\mathrm{N2} = \log(\mathrm{[N\,\textsc{ii}]6584}/{\mathrm{H}\alpha})$) and the corresponding abundance calibration \citep{kd02, denicolo02, pmc09}.  

The aim of this paper is to investigate \---~for the first time at a redshift around 1 \--- metallicity gradients of 50 star-forming galaxies in the MASSIV sample. In order to achieve this goal, we develop a dedicated program to analyse the low SNR data cubes. Two companion papers \citep{vergani11, epinat11} discuss the associated fundamental scaling relations (e.g.~the Tully-Fischer relation for disks) and the kinematical properties of the sample galaxies.  
The results of these three studies are cross-correlated in order to investigate possible relations between the metallicity gradients and  global properties of galaxies at $z\sim1.2$.  

The paper is organized as follows. In Section~\ref{sec:datadescr} we briefly summarise the properties of the galaxy sample, the data reduction technique, the emission-line measurements, the level of Active Galactic Nucleus (AGN) contamination, the kinematics, and the mass (both stellar and dark matter halo) estimates. In Section~\ref{sec:zgrad} we present and discuss the determined metallicity gradients. Our conclusions are drawn in Section~\ref{sec:conclu}.

Throughout the paper, we assume a $\Lambda$CDM cosmology with $\Omega_m=0.3$, $\Omega_\Lambda=0.7$ and $H_0 = 70~ \mathrm{km s}^{-1}\mathrm{ Mpc}^{-1} $.

\section{Data Description}
\label{sec:datadescr}

\subsection{Target selection, observations and data reduction}

The galaxy sample studied in this paper is the ``first epoch'' sample of the MASSIV project (ESO Large Program, PI.: T. Contini). A full description of the sample can be found in \cite{contini11}. 
We briefly summarise some properties of this sample of 50 galaxies below.

The galaxies were selected from the VIMOS VLT Deep Survey (VVDS) in the deep \citep[$I_{AB}<24$;][]{lefevre05}, ultra-deep
($I_{AB}<24.75$; Le F\`evre et al., in prep ) $\mathrm{RA} = 02\mathrm{h}$, and wide \citep[$I_{AB}<22.5$;][]{garilli08} $\mathrm{RA}=14\mathrm{h}, 22\mathrm{h}$ fields. The galaxies were chosen according to their redshift such that their \ha{} line (or [O\,\textsc{iii}]5007 in a few cases) was visible in the $J$ or $H$ band, and was not affected by a bright OH sky-line.  Galaxies were selected to be star-forming on the basis of their [O\,\textsc{ii}]3727 emission line strength.  The observations have been performed between April 2007 and August 2009. The general properties (RA, DEC, redshift, $I$-band magnitude) and characteristic observational configuration (spectral band, adaptive optics or not, exposure time, spatial resolution) of each galaxy are gathered in Table~\ref{gene}. Most ($85$\%) of the galaxies in this ``first epoch'' sample have been observed in a seeing-limited mode (with a spatial sampling of $0.125$\arcsec{}). However, seven galaxies have been acquired with Adaptive Optics assisted with a Laser Guide Star (AO/LGS, $0.05$\arcsec{} spatial sampling).  Among the 50 ``first epoch" MASSIV galaxies observed with SINFONI, four galaxies have not been detected \citep[see][for details]{epinat11}.

The data reduction was performed with the ESO SINFONI pipeline, using the standard master calibration files provided by ESO. The absolute astrometry for the SINFONI data cubes was derived from nearby bright stars also used for PSF measurements. Custom \textsf{IDL} and \textsf{Python} scripts have been used to flux calibrate, align, and combine all the individual exposures. For each galaxy a non sky-subtracted cube was also created, mainly to estimate the effective spectral resolution. For more details on data reduction, we refer to \cite{epinat11}.

$I$-band images for all the galaxies were obtained through CFHT Megacam imaging (for the 22h and
02h field, from the CFHTLS ``best seeing'') and CFHT-12K imaging (for the 14h field).  These images were used for two purposes: (i) refinement of the SINFONI astrometry, using the relative position of the PSF star, and (ii), deriving morphological parameters \citep[used as inputs to the kinematics modeling, see][]{epinat11}.

\begin{table*}
    \caption{General properties and SINFONI observing log of the 50 ``first
      epoch'' galaxies of MASSIV. Column~1: identifier in the VVDS catalog, 
      Columns~2 \& 3: RA \& DEC in
      degrees, Column~4: spectral band of the observation, Column~5:
      seeing-limited mode (0) or Adaptive Optics (1), Column~6: on-source
      total exposure time, Column~7: $I$-band magnitude in the AB system,
      Column~8: Mean spatial resolution (PSF) as measured on a nearby
      star, Column~9: redshift from \ha{} (or [O\,\textsc{iii}]5007 when marked with 
      an asterisk) as measured with SINFONI. For the four galaxies (VVDS020126402, 
      VVDS020217890, VVDS020306817, and VVDS220071601) which have not been 
      detected with SINFONI, redshift comes from VIMOS spectra}\label{gene}
  \begin{center}
    \begin{tabular}{ccccccccc}
      \hline
      \hline
      Galaxy & RA (J2000) & Dec (J2000) & Band & AO/LGS &$t_\mathrm{exp}$ & $I_\mathrm{AB}$ & Mean PSF & $z_\textrm
{SINFONI}$ \\
      & [deg] & [deg] & & & [sec] & [mag] & [\arcsec] & \\ 
      (1) & (2) & (3) & (4) & (5) & (6) & (7) & (8) & (9) \\
      \hline
      020106882 & 36.340833 & -4.771834 & H & 0 & 4800 & 23.18 & 0.49 & 1.3991 \\
      020116027 & 36.463055 & -4.751243 & H & 0 & 4500 & 22.88 & 0.60 & 1.5302 \\
      020126402 & 36.298576 & -4.727812 & J & 1 & 3600 & 22.99 & $\dots$ & 1.2332 \\
      020147106 & 36.689110 & -4.679830 & H & 0 & 7200 & 22.51 & 0.65 & 1.5195 \\
      020149061 & 36.771769 & -4.674816 & H & 0 & 4800 & 22.56 & 0.85 & 1.2905 \\
      020164388 & 36.712244 & -4.639131 & H & 0 & 4800 & 22.51 & 0.83 & 1.3547 \\
      020167131 & 36.697114 & -4.632054 & J & 0 & 7200 & 23.01 & 0.68 & 1.2246$^*$ \\
      020182331 & 36.684314 & -4.597755 & H & 0 & 10800 & 22.73 & 0.74 & 1.2290 \\
      020193070 & 36.327984 & -4.572212 & J & 0 & 7200 & 23.41 & 0.58 & 1.0279 \\
      020208482 & 36.319735 & -4.536649 & J & 0 & 7200 & 23.25 & 0.58 & 1.0375 \\
      020214655 & 36.597688 & -4.523057 & J & 0 & 4800 & 23.05 & 0.87 & 1.0395 \\
      020217890 & 36.613174 & -4.514397 & H & 0 & 7200 & 23.99 & $\dots$ & 1.5129 \\
      020239133 & 36.679087 & -4.475228 & J & 0 & 4800 & 22.85 & 0.79 & 1.0194 \\
      020240675 & 36.725581 & -4.471541 & H & 0 & 4800 & 23.45 & 0.85 & 1.3270 \\
      020255799 & 36.691075 & -4.437713 & J & 0 & 4800 & 23.63 & 0.76 & 1.0351 \\
      020261328 & 36.796038 & -4.425444 & H & 0 & 3600 & 23.90 & 0.62 & 1.5290 \\
      020278667 & 36.492598 & -4.386738 & J & 0 & 7200 & 23.25 & 0.65 & 1.0516 \\
      020283083 & 36.628634 & -4.376830 & H & 0 & 4800 & 23.07 & 0.78 & 1.2818 \\
      020283830 & 36.620565 & -4.375444 & H & 0 &7200 & 22.92 & 0.77 & 1.3949 \\
      020294045 & 36.446413 & -4.352110 & J & 0 & 7200 & 22.80 & 0.59 & 1.0028 \\
      020306817 & 36.459649 & -4.323037 & J & 0 & 7200 & 23.29 & $\dots$ & 1.2225 \\
      020363717 & 36.598723 & -4.199505 & H & 0 & 4800 & 22.61 & 0.64 & 1.3339 \\
      020370467 & 36.561244 & -4.184841 & H & 0 & 4800 & 23.46 & 0.71 & 1.3338 \\
      020386743 & 36.808285 & -4.149879 & J & 0 & 7200 & 22.58 & 0.73 & 1.0487 \\
      020461235 & 36.696290 & -4.398832 & J & 0 & 7200 & 22.64 & 0.63 & 1.0349 \\
      020461893 & 36.801068 & -4.386450 & J & 0 & 4800 & 23.45 & 0.60 & 1.0486 \\
      020465775 & 36.747379 & -4.316665 & H & 0 & 4800 & 23.15 & 0.88 & 1.3583 \\
      140083410 & 209.460955 & 4.294251 & J & 0 & 4800 & 21.82 & 0.69 & 0.9435 \\
      140096645 & 209.609695 & 4.329940 & J & 0 & 7200 & 22.28 & 0.56 & 0.9655 \\
      140123568 & 208.990111 & 4.405582 & J & 0 & 7200 & 23.43 & 0.76 & 1.0012 \\
      140137235 & 209.053257 & 4.442166 & J & 0 & 4800 & 22.38 & 0.76 & 1.0445 \\
      140217425 & 209.485039 & 4.643635 & J & 0 & 6000 & 21.58 & 0.95 & 0.9792 \\
      140258511 & 210.081944 & 4.746065 & H & 0 & 4800 & 21.17 & 0.49 & 1.2423 \\
      140262766 & 209.981154 & 4.758375 & H & 0 & 7200 & 23.68 & 0.51 & 1.2836 \\
      140545062 & 209.898275 & 5.508636 & J & 0 & 7200 & 22.42 & 0.70 & 1.0408 \\
      220014252 & 334.440374 & 0.477630 & H & 0 & 7200 & 22.04 & 0.70 & 1.3105 \\
      220015726 & 333.926894 & 0.484332 & H & 0 & 7200 & 22.42 & 0.46 & 1.2933 \\
      220071601 & 334.506537 & 0.759637 & H & 1 & 4800 & 21.74 &  $\dots$ & 1.3538 \\
      220148046 & 333.657514 & 1.139144 & H & 1 & 4800 & 22.39 & 0.27 & 2.2441$^*$ \\
      220376206 & 335.024084 & -0.139356 & H & 0 & 7200 & 21.78 & 0.50 & 1.2445 \\
      220386469 & 334.985762 & -0.050878 & J & 1 & 2400 & 22.10 & 0.23 & 1.0226 \\
      220397579 & 335.152164 & 0.029634 & J & 0 & 7200 & 22.42 & 0.64 & 1.0379 \\
      220544103 & 333.857118 & 0.110982 & H & 0 & 7200 & 22.38 & 0.76 & 1.3973 \\
      220544394 & 333.600634 & 0.113027 & J & 0 & 7200 & 22.15 & 0.58 & 1.0101 \\
      220576226 & 334.047658 & 0.275121 & J & 0 & 7200 & 21.80 & 0.58 & 1.0217 \\
      220578040 & 334.267094 & 0.282327 & J & 0 & 7200 & 22.30 & 0.62 & 1.0462 \\
      220584167 & 333.845992 & 0.313059 & H & 0 & 7200 & 21.96 & 0.75 & 1.4655 \\
      220596913 & 333.621601 & 0.371914 & H & 1 & 7200 & 21.80 & 0.18 & 1.2658 \\
      910193711 & 36.442825 & -4.542796 & H & 1 & 4800 & 22.69 & 0.27 & 1.5564 \\
      910279515 & 36.401024 & -4.354381 & H & 1 & 4800 & 23.71 & 0.21 & 1.4013 \\
      \hline
    \end{tabular}
  \end{center}
\end{table*}

\subsection{Measurement of emission lines}\label{el}

As mentioned in the previous section, galaxies were selected to be star-forming such that emission lines from H\,\textsc{ii} regions are visible in the near infrared. Given the redshift range of MASSIV galaxies, we targeted, in most of the cases, the rest-frame optical \ha{} emission line in the $J$ (for $z <1.2$) or $H$ (for $z>1.2$) band.  For four objects (VVDS020126402, 020167131, 020306817, and 220148046) we observed the [O\,\textsc{iii}]5007 line which allowed to increase the redshift window  (two of them were not detected). These latter galaxies are not considered in this paper as we cannot derive any metallicity from the [O\,\textsc{iii}]5007 emission-line alone.  In the former cases, we took advantage of the proximity of the nitrogen line to \ha{} to derive the metallicity of a majority of the galaxies using the \niia{}/\ha{} line ratio as a proxy for oxygen abundance (see sect.~\ref{sec:zgrad}).

\subsubsection{Integrated fluxes}\label{intfl}
The global integrated flux of each emission line was calculated using a mask designed for each galaxy.  The flux was collected over a region where the signal-to-noise ratio of the \ha{} line is above a fixed threshold of $SNR>2$.  The SNR maps were produced using a Gaussian-fitting procedure written in \textsf{IDL}, already used and described in \cite{epinat11}.  A Gaussian spatial smoothing of $2\times 2$ spaxels (0.25\arcsec$\times$ 0.25\arcsec) was used, increasing the SNR without degrading the final spatial resolution ($\sim 0.65$\arcsec\ on average).  The line position map produced by this procedure was used to shift in wavelength every single spectrum such as to compensate for the Doppler effect corresponding to the measured \ha{} velocity at that position. In this way, the broadening of the line by large-scale rotation in the spatially-integrated spectrum was largely removed and the line profile more easily fitted with a Gaussian, increasing the accuracy of line measurements. 

The \ha{} fluxes were measured on flux-calibrated integrated spectra using a Gaussian fit and a flat continuum. These fluxes have been corrected
for dust reddening using the extinction coefficient derived from the SED fitting \citep[see][]{contini11, vergani11}. The line ratios, on the other hand, were determined from the integrated spectra in counts. Indeed, the flux calibration procedure would have added noise on top of the already low SNR data. Further, the lines of interest (namely, \ha{} and \niia) are close enough in wavelength to assume that the sensitivity curve is constant over the wavelength range of interest. The same argument justifies the fact that we did not correct the emission-line ratios for differential extinction.  Table~\ref{eml} lists our final integrated \ha{} flux, N2 emission line ratio 
($N2 = \log(\mathrm{\niia{}}/\mathrm{H\alpha})$ and star formation rates (SFR, corrected or not for dust reddening).  Among the 44 galaxies detected in \ha{} with SINFONI, we have been able to measure the \niia{} emission-line in 34 galaxies and hence derive integrated N2 emission-line ratio and metallicity for these objects.  

\tabcolsep1.40mm
\begin{table}[!ht]
  \caption{\ha{} flux, $N2= \log(\mathrm{\niia{}}/\mathrm{H\alpha}$) emission-line ratio, and \ha{}-based star formation rates 
(corrected \--- and not \--- for dust reddening) for each galaxies in our sample.}\label{eml}
  \centering
  \begin{tabular}{crrrr}
    \hline
    \hline
    Galaxy & F(H$\alpha$) & N2 & $SFR_{\rm{H}\alpha}$ & $SFR_{\rm{H}\alpha}^{\rm cor}$  \\
    & [${10}^{-17}$ergs\un{}cm\car{}] & & [\msun{} yr\un{}] &  [\msun{} yr\un{}] \\
    (1) & (2) & (3) & (4) & (5) \\
    \hline
    020106882 & $14.0 \pm  0.8$ & $-0.71 \pm 0.12$ & 13.3 & 38.1 \\
    020116027 & $12.4 \pm  2.9$ & $-1.02 \pm 0.15$ & 14.7 & 42.2 \\
    020126402 & $\dots$ & $\dots$ & $\dots$ & $\dots$ \\
    020147106 & $46.8 \pm  5.8$ & $-0.95 \pm 0.14$ & 54.3 & 92.1 \\
    020149061 & $20.6 \pm  1.6$ & $\dots$ & 16.0 & 45.9 \\
    020164388 & $24.1 \pm  1.3$ & $-0.86 \pm 0.10$ & 21.1 & 46.5 \\
    020167131 & $\dots$ & $\dots$ & $\dots$ & $\dots$ \\
    020182331 & $15.2 \pm  5.0$ & $\dots$ & 10.5 & 51.1 \\
    020193070 & $ 6.7 \pm  0.9$ & $-0.56 \pm 0.13$ & 3.0 & 14.4 \\
    020208482 & $ 2.6 \pm  0.5$ & $\dots$ & 1.2  & 2.6 \\
    020214655 & $23.3 \pm  1.1$ & $-0.68 \pm 0.06$ & 10.6 & 51.7 \\
    020217890 & $\dots$ & $\dots$ & $\dots$ & $\dots$ \\
    020239133 & $ 7.8 \pm  1.0$ & $\dots$ & 3.4 & 16.5 \\
    020240675 & $ 8.5 \pm  1.0$ & $\dots$ & 7.1 & 15.7 \\
    020255799 & $ 5.8 \pm  0.7$ & $-0.65 \pm 0.15$ & 2.6 & 12.8 \\
    020261328 & $ 9.5 \pm  0.9$ & $-1.06 \pm 0.48$ & 11.2 & 19.0 \\
    020278667 & $ 2.4 \pm  1.3$ & $\dots$ & 1.1  & 4.2 \\
    020283083 & $13.4 \pm  1.0$ & $-0.70 \pm 0.12$ & 10.3 & 17.4 \\
    020283830 & $11.0 \pm  1.1$ & $-0.82 \pm 0.17$ & 10.3 & 50.4 \\
    020294045 & $15.5 \pm  2.8$ & $\dots$ & 6.5 & 14.3 \\
    020306817 & $\dots$ & $\dots$ & $\dots$ & $\dots$ \\
    020363717 & $37.4 \pm  4.3$ & $-0.87 \pm 0.09$ & 31.5 & 53.4 \\
    020370467 & $17.3 \pm  2.9$ & $-0.69 \pm 0.11$ & 14.6 & 71.2 \\
    020386743 & $17.4 \pm  1.7$ & $-0.91 \pm 0.08$ & 8.1 & 39.4 \\
    020461235 & $ 7.5 \pm  2.0$ & $-0.67 \pm 0.12$ & 3.3 & 9.6 \\
    020461893 & $14.5 \pm  1.2$ & $\dots$ & 6.7 & 14.9 \\
    020465775 & $14.5 \pm  2.8$ & $\dots$ & 12.8 & 62.2 \\
    140083410 & $47.0 \pm  3.3$ & $-0.72 \pm 0.08$ & 16.8 & 37.1 \\
    140096645 & $55.4 \pm  4.2$ & $-0.25 \pm 0.02$ & 20.9 & 102.1 \\
    140123568 & $ 5.8 \pm  1.2$ & $-0.69 \pm 0.10$ & 2.4 & 11.8 \\
    140137235 & $ 9.6 \pm  2.3$ & $-0.79 \pm 0.13$ & 4.4 & 21.6 \\
    140217425 & $104.7 \pm  4.9$ & $-0.44 \pm 0.02$ & 41.0 & 200.0 \\
    140258511 & $42.5 \pm  2.7$ & $-0.51 \pm 0.10$ & 30.0 & 146.4 \\
    140262766 & $13.1 \pm  5.0$ & $-0.75 \pm 0.15$ &  10.0 &  10.0 \\
    140545062 & $29.4 \pm  3.3$ & $-0.75 \pm 0.06$ & 13.4 & 22.7 \\
    220014252 & $50.3 \pm  3.1$ & $-0.67 \pm 0.08$ & 40.5 & 197.5 \\
    220015726 & $47.5 \pm  6.7$ & $-0.69 \pm 0.04$ & 37.0 & 106.4 \\
    220071601 & $\dots$ & $\dots$ & $\dots$ & $\dots$ \\
    220148046 & $\dots$ & $\dots$ & $\dots$ & $\dots$ \\
    220376206 & $72.4 \pm  5.1$ & $-1.04 \pm 0.10$ & 51.2 & 249.7 \\
    220386469 & $17.1 \pm  2.4$ & $-1.14 \pm 0.16$ & 7.5 & 36.5 \\
    220397579 & $65.1 \pm  9.1$ & $-1.07 \pm 0.07$ & 29.4 & 143.2 \\
    220544103 & $55.0 \pm  1.9$ & $-0.56 \pm 0.09$ & 52.2 & 117.5 \\
    220544394 & $24.3 \pm  1.7$ & $-0.89 \pm 0.07$ & 10.3 & 50.1 \\
    220576226 & $31.2 \pm  1.6$ & $-0.69 \pm 0.04$ & 13.6 & 66.3 \\
    220578040 & $20.5 \pm  1.9$ & $-0.80 \pm 0.07$ & 9.4 & 20.8 \\
    220584167 & $64.3 \pm  1.3$ & $-0.91 \pm 0.10$ & 68.6 & 202.6 \\
    220596913 & $30.0 \pm  1.0$ & $\dots$ &  23.2 & 36.1 \\
    910193711 & $32.9 \pm 18.4$ & $-0.68 \pm 0.09$ & 40.6 & 197.7 \\
    910279515 & $14.9 \pm  2.1$ & $-0.83 \pm 0.21$ & 14.2 & 69.0 \\
    \hline
  \end{tabular}
\end{table}

\subsubsection{Spatially-resolved flux and line ratio}

For many galaxies in our sample, the low SNR of the data cubes prevents us from measuring in each spaxel, with high confidence, the emission lines surrounding \ha{} : namely \niia, \sii{}, and [O\,\textsc{i}]6300 which is detected only in one galaxy: VVDS140096645.  
These lines might either not be bright enough to be detected, or a bright OH sky-line is too close.  To
increase the SNR of our final spectra, we summed the spectra of the spaxels gathered in a specified region, and measured the line fluxes in the resulting spectra.  

\begin{figure}[t]
  \centering
  \raisebox{0.7cm}{\includegraphics[width=0.5\linewidth]{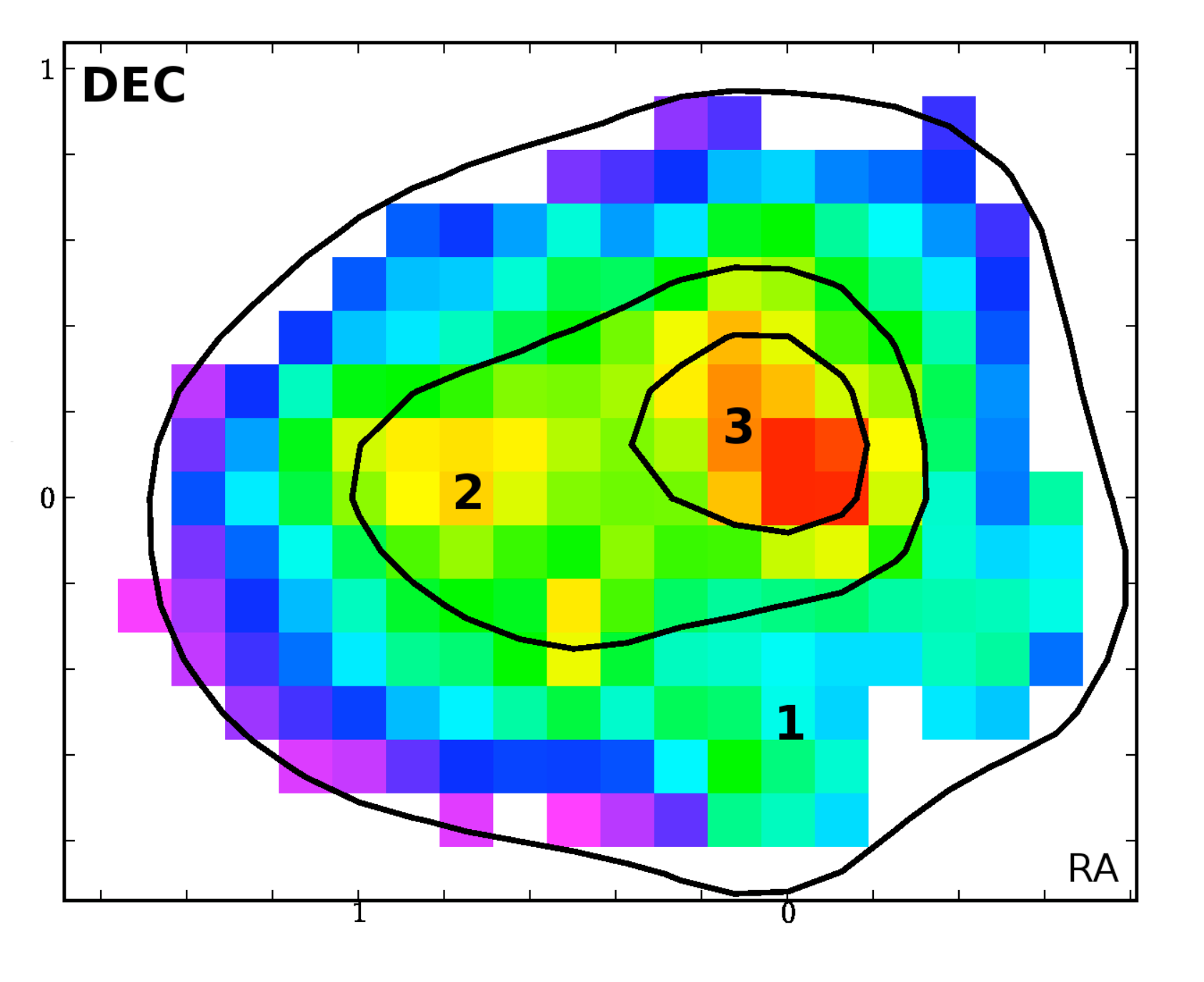}}\includegraphics[width=0.5\linewidth]{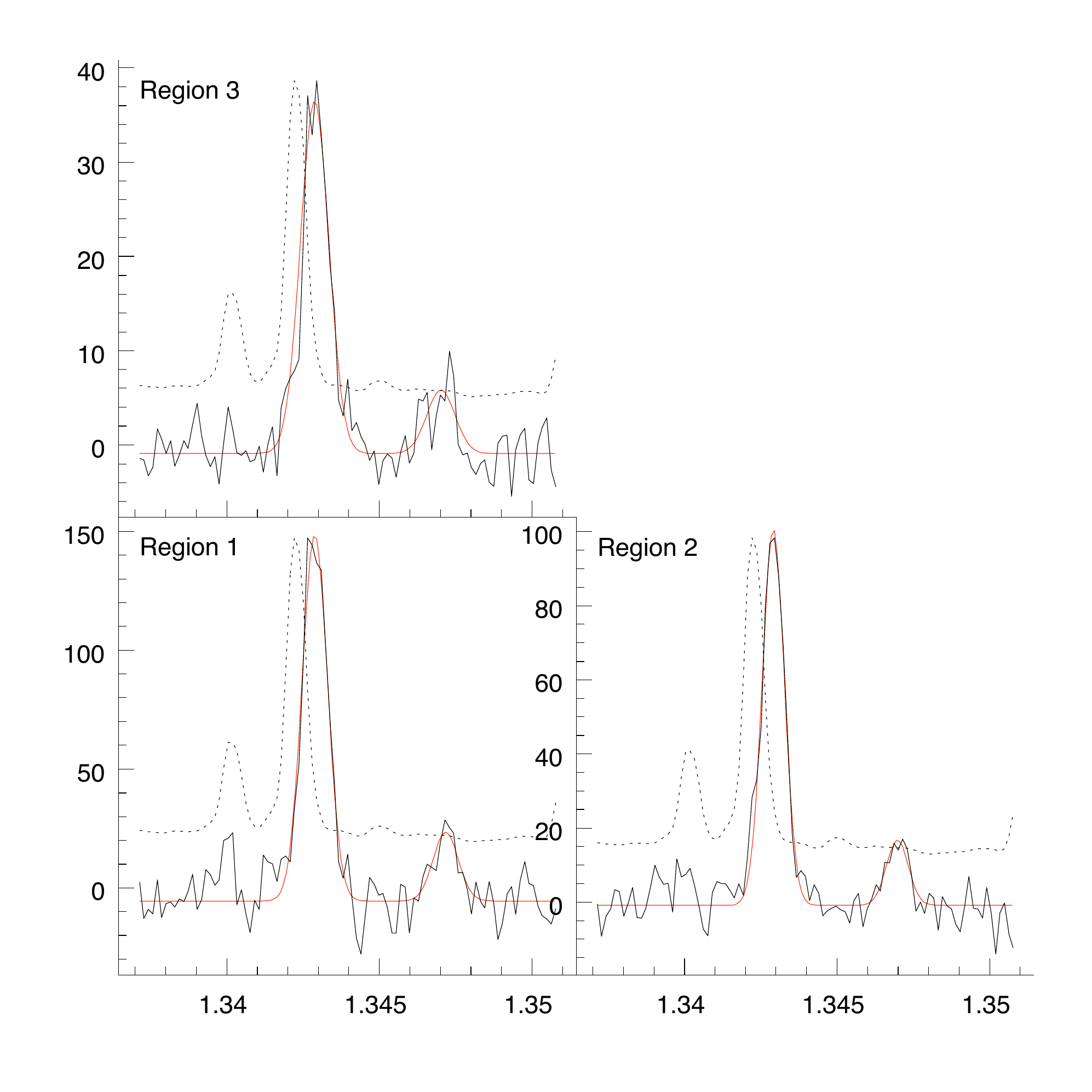}

  \includegraphics[width=0.5\linewidth]{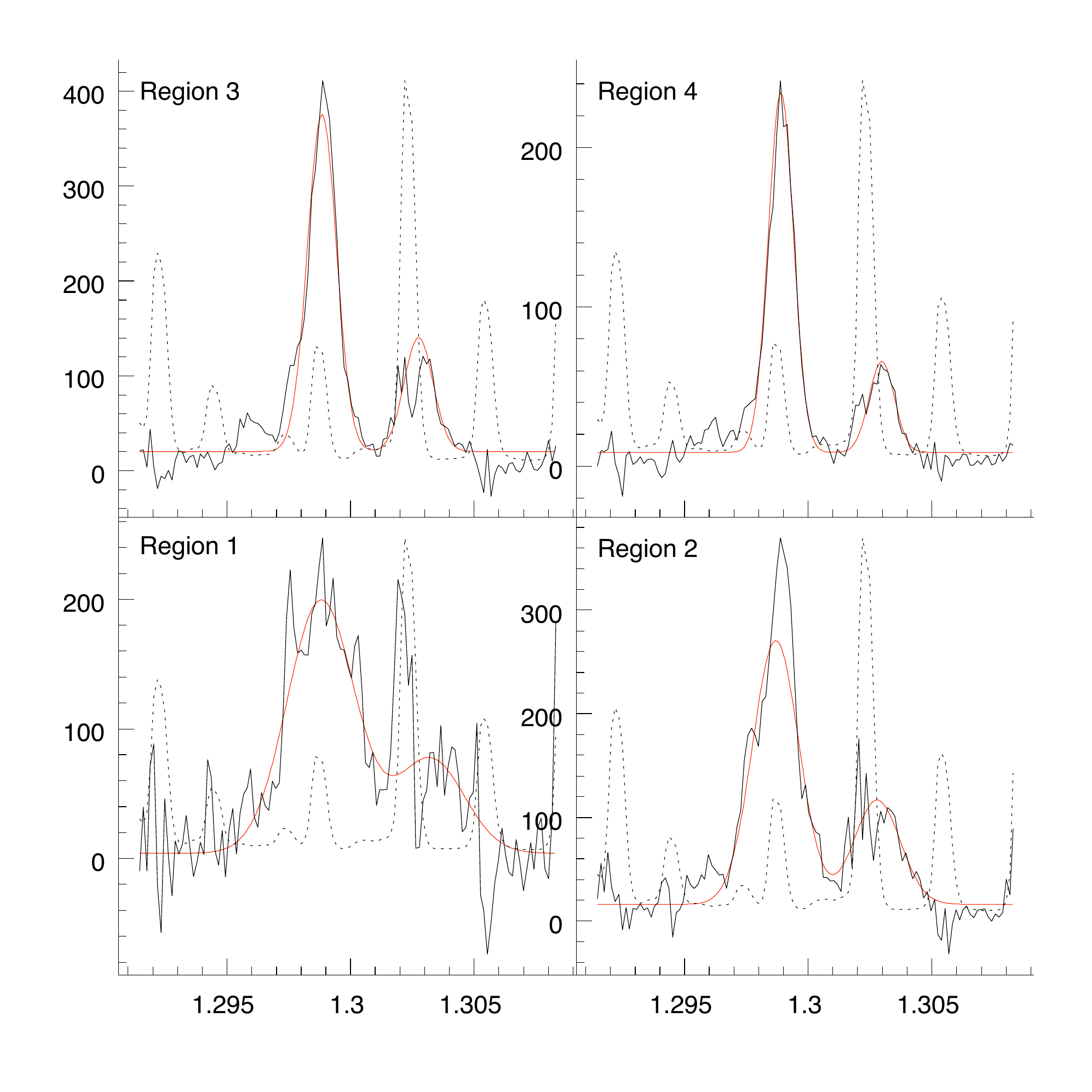}\raisebox{0.7cm}{\includegraphics[width=0.5\linewidth]{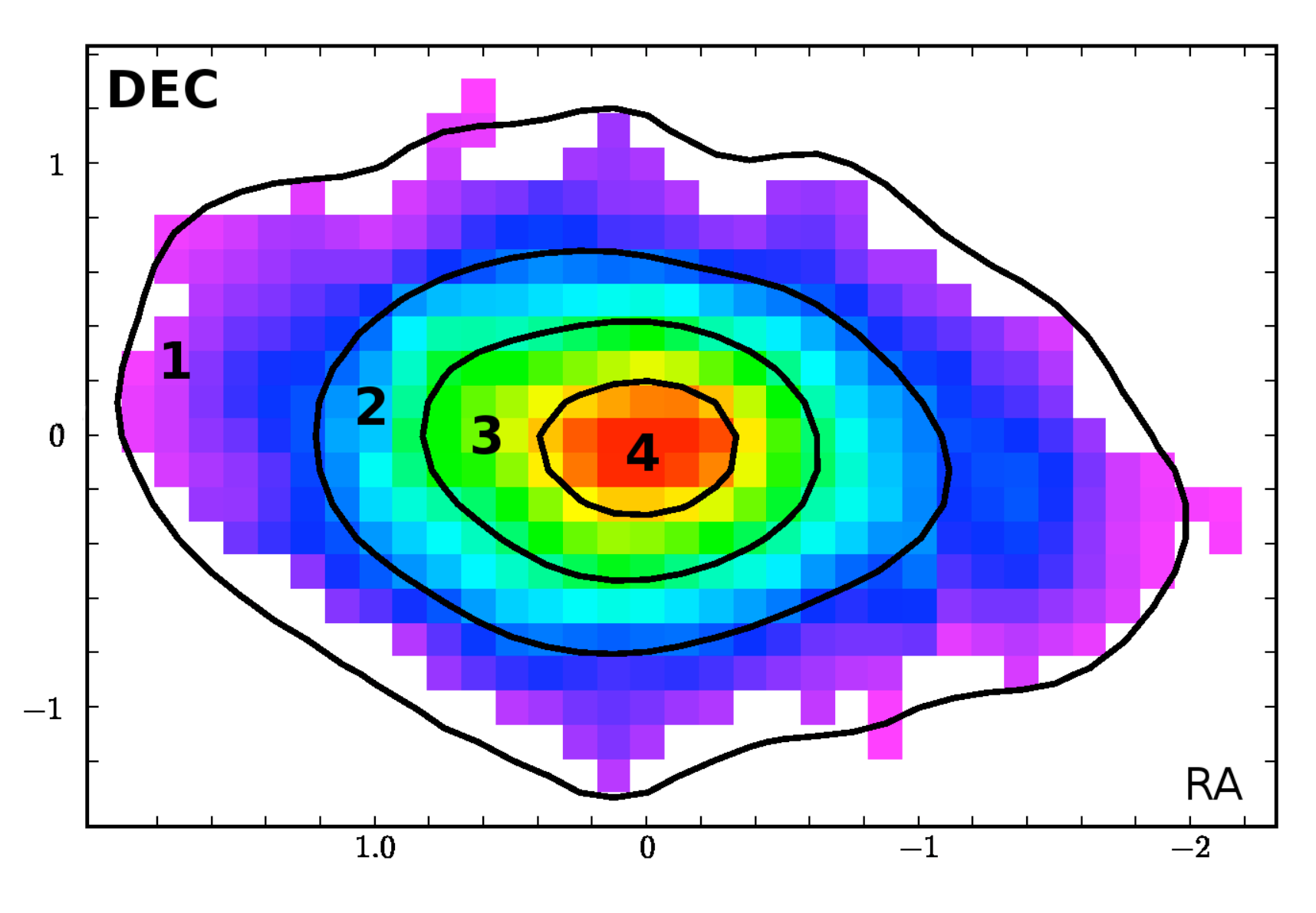}}
\caption{Examples
    of spatially-resolved emission-lines measurements in two MASSIV
    galaxies (Top: VVDS220578040, bottom: VVDS140217425). Color-coded
    images show the regions defined with \ha{} contours used to derive
    the metallicity gradients. The 1D spectra integrated over the
    different regions are shown with solid black lines. Model (in red)
    and sky (dashed line) spectra are also shown. The sky spectrum is
    used as a weight in the fitting procedure. Region 1 is the outer most with the region number increasing towards the center.}\label{prog}
\end{figure}

We developed a specific \textsf{Python} procedure to do this analysis within our data cubes.
The program allowed to define a region spaxel-by-spaxel, or to define a region with contours on 2D maps (such as \ha{} flux, SNR, etc maps). In a given region, the spaxel-to-spaxel line shift due to the rotation velocity of the gas along the line-of-sight was fully corrected for (see \S\ref{intfl}). The integrated spectrum in a defined region was then fitted with a flat continuum and two Gaussians, one for \ha{} and one for \niia{} (the sulfur doublet was neglected for this analysis as it was detected in a few cases only).  The width of each Gaussian was set to be the same as the emission of the collisional and recombination lines traces the same ionised gas in the galaxies.  The
fit was weighted using the corresponding non sky-subtracted spectrum, i.e.~giving a lower weight to the channels affected by the subtraction of a strong sky-line.  We then estimated the $1\sigma$ error on the fluxes with a Monte-Carlo technique, in which the fit was considered to be a ``noise-free'' model, to which we added a Gaussian noise with a standard deviation corresponding to the residuals from the initial spectrum minus the ``noise-free'' model.  This operation was repeated a hundred of times, which gave a set of parameters on which the $1\sigma$ deviation was computed.  
\fig{prog} shows a result of our fits in different regions for two galaxies.

Finally, we estimate spatially-resolved \ha\ and \niia\ emission-line flux for 26/34 ($\sim 75$\%) galaxies in the sample. The remaining 8 galaxies have either a too low SNR or 
the \niia\ emission-line is polluted by sky-line residuals.

\subsection{AGN contamination}

A recent study \citep{wright10} has shown that unveiling the presence of AGN in high-redshift galaxies is a difficult exercise.  In the case of metallicity studies, in which abundances are deduced from the ratio of
different emission lines of the ionised gas, it is critical to check that the intensity and width of these lines are due to star formation and not related to any non-thermal nuclear activity.  The common way to disentangle AGN contribution from star-forming galaxies consists in comparing the relative intensity of the main nebular
emission lines (mainly [O\,\textsc{iii}]5007, H$\beta$, H$\alpha$, and \niia{}) in a diagnostic diagram, so-called BPT diagram \citep{bpt81, kewley01}.  Various physical conditions in the ISM \---  SFR,  ionisation parameter, metallicity and/or chemical composition \--- have been invoked to explain the fact that some high-$z$ star-forming galaxies lie in the transition region of the local BPT diagram (as defined by the SDSS galaxies) between
star-forming galaxies and AGN hosts.  \citet{wright10} have been able \--- thanks to high resolution adaptive optics observations \--- to subtract the active nuclear emission in a $z\sim 1.6$ galaxy (HDF-BMZ1299), and have shown that the residual extended star-forming emission was characteristic of a local SDSS star-forming galaxy, whereas the integrated emission would have placed the object in the transition region.
\begin{figure}[!ht]
  \centering
  \includegraphics[width=0.5\linewidth]{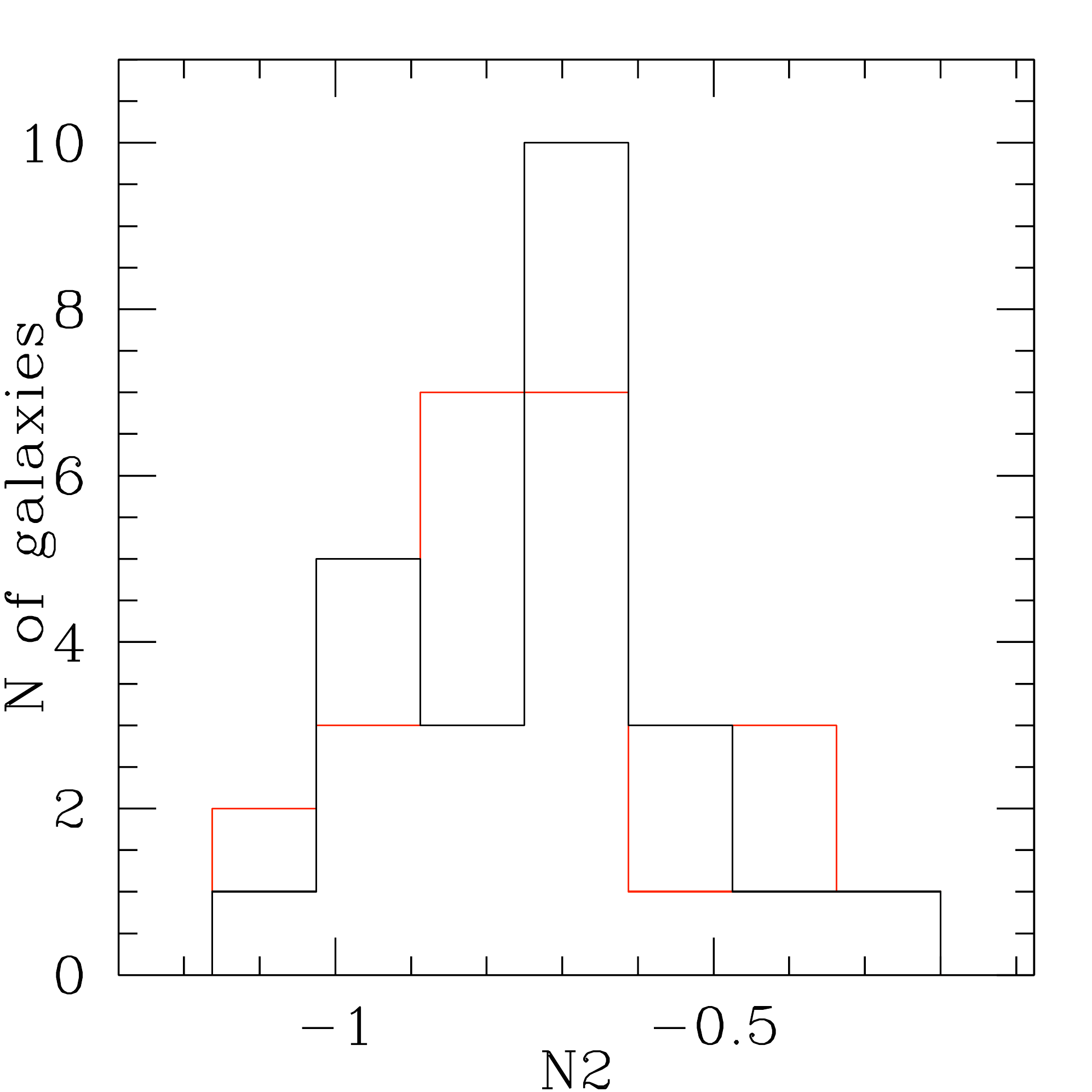}\includegraphics[width=0.5\linewidth]{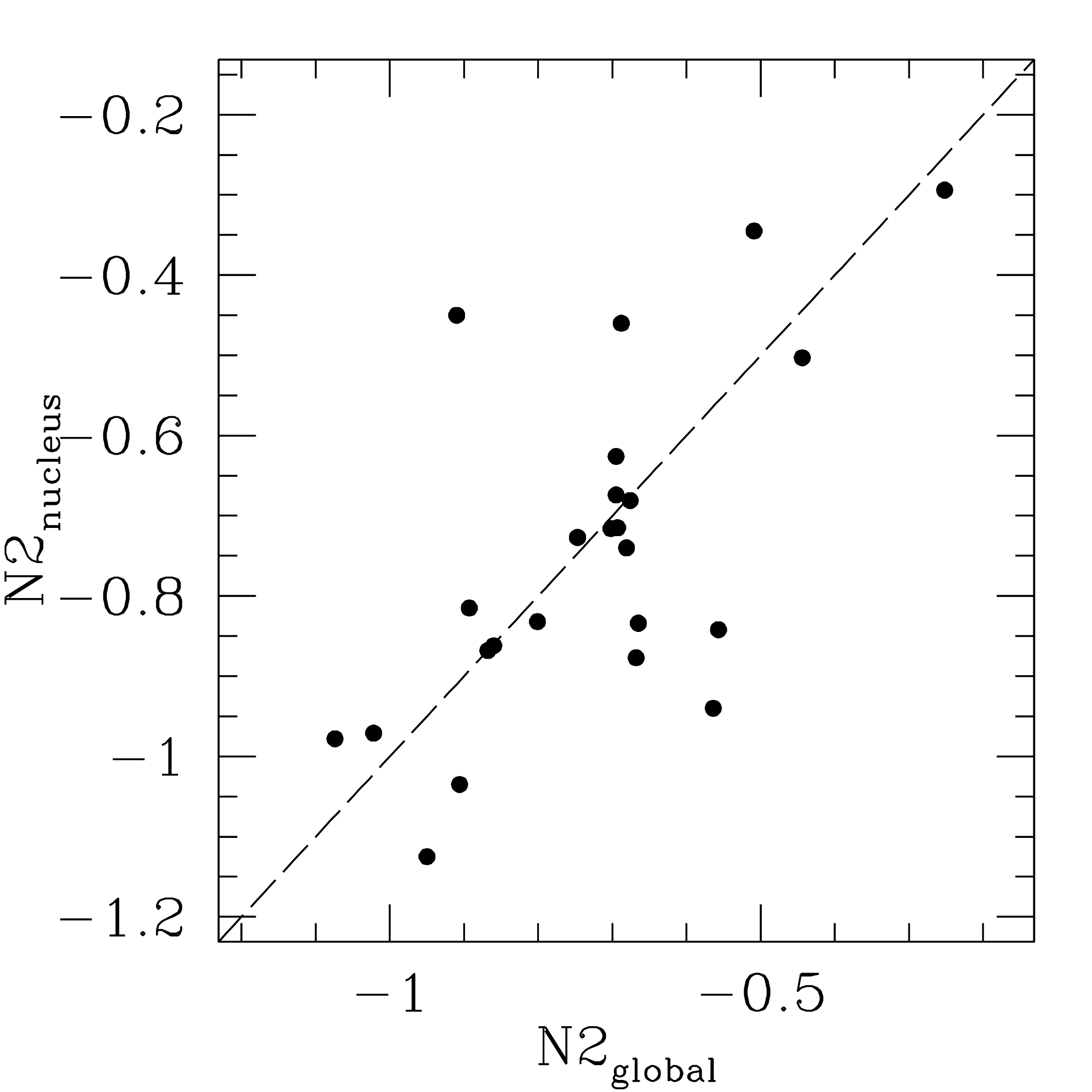}
  \caption{{\it Left}: Distribution of the N2 ratio in the MASSIV ``first
    epoch'' sample. The black line represents the ratio integrated over
    the whole galaxy. The red line is the N2 ratio measured in the
    nuclear region. {\it Right}: N2 global ratio versus N2 nuclear
    ratio.}\label{distrib}
\end{figure}

The nature of our observations did not give us simultaneous access to the set of emission lines
([O\,\textsc{iii}]5007, H$\beta$, H$\alpha$, and [N\,\textsc{ii}]6584 or \sii{}) commonly used in standard diagnostic diagrams. However, for all but two objects in our sample, the emission-line ratio $\mathrm{N2} =
\log(\textrm\niia{}/\textrm\ha)$ is lower than $-0.5$, with a median $\mathrm{N2}$ value of $-0.72$. Such low values are indicative for a very low contamination by AGN in our sample \citep[eg.][]{bpt81}. 

For 24 galaxies of our sample we  calculated, following \citet{wright10}, the N2 ``concentrated ratio'', corresponding to the value in the nuclear region of the galaxy. We defined the nuclear region as the spaxel with the highest \ha{} flux along with its 8 nearest neighbours (corresponding to a $0.7''$ diameter aperture, matching our mean spatial resolution of $0.65''$).  We assumed this aperture to be small enough to probe the inner
nucleus part as objects usually span from $1''$ to $2''$ with $SNR>2$ in our observations. 

\begin{figure}[t]
  \centering
  \includegraphics[width=\linewidth]{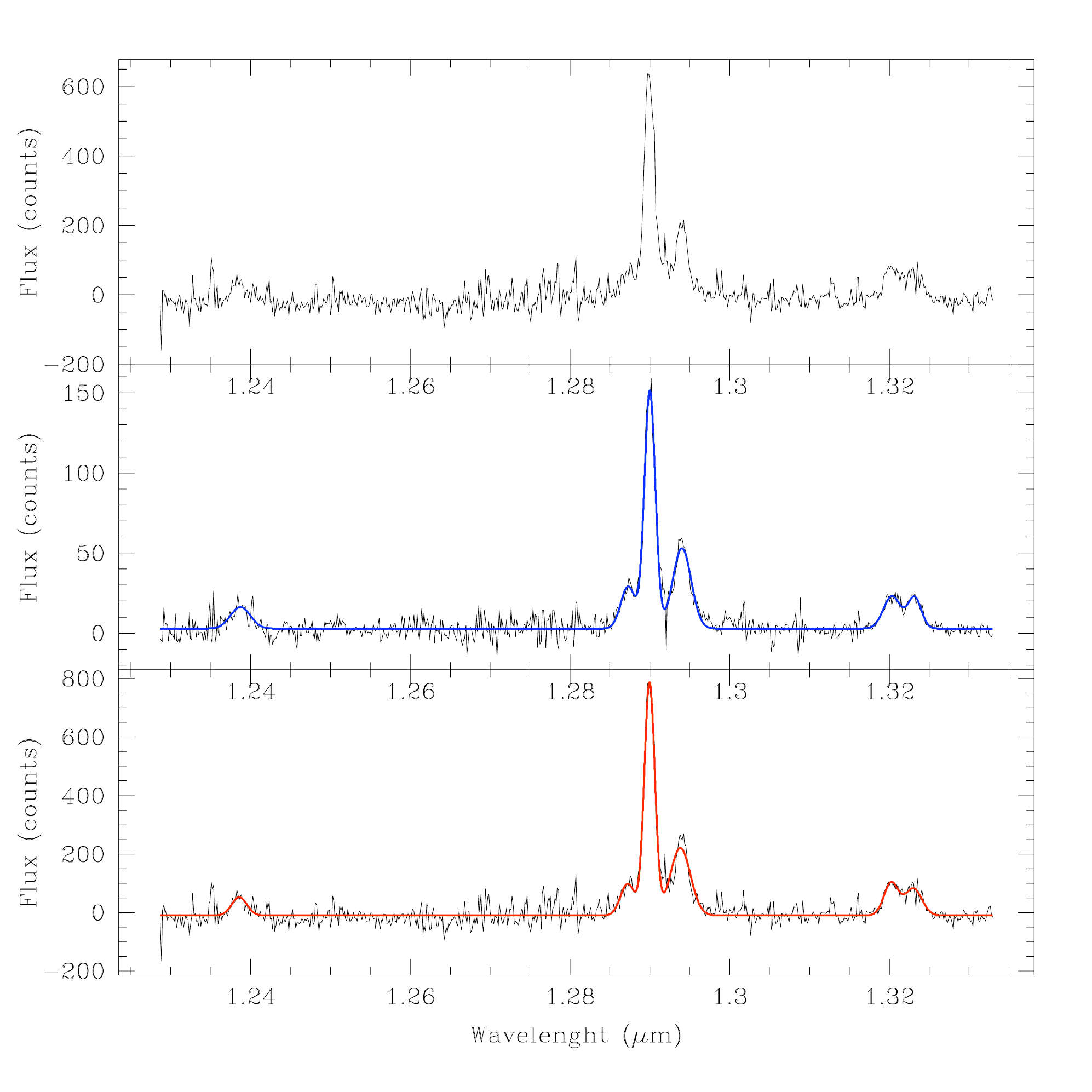}
  \caption{Integrated spectra of VVDS140096645. {\it Top}: spectrum of the
    disk (total integrated with central contribution removed). {\it Middle}:
    spectrum of the central region surrounding the peak of \ha{}
    flux. {\it Bottom}: spectrum from the whole spatial extent. From left to
    right, the following emission lines appear: [\textsc{O\,i}]6300,
    [\textsc{N\,ii}]6548, \ha{}, [\textsc{N\,ii}]6584,
    [\textsc{S\,ii}]6717, [\textsc{S\,ii}]6731. The red and blue lines
    are least-square fits to the data.}\label{645}
\end{figure}

Fig.~\ref{distrib} shows i) the distribution of the 24 galaxies as a function of their global and nuclear N2 ratio (left panel) and ii) the relation between the global and nuclear N2 ratios for each galaxy (right panel). The median values of N2 for each distribution are not very different ($\Delta\sim -0.07$). The median nuclear N2 ratio is lower than the global ratio which would not be the case if a significant fraction of our sample galaxies were hosting an AGN. When comparing the global ratio to the nucleus ratio distribution, the highest bin does not shift and contains a single object . We investigated in more detail the galaxy in this bin: VVDS140096645. It shows the following high N2 ratios: $N2_\textrm{global}=-0.252$ and $N2_\textrm{nuclear}=-0.294$.  Looking further into its integrated spectrum (global and nuclear, see \fig{645}), we noticed that: (i) the emission lines are broad, which is a possible sign of nuclear activity, (ii) the two nitrogen lines are clearly visible, as is the sulfur doublet, and the [O\,\textsc{i}]6300 line, which altogether are characteristics of LINER galaxies, often associated to AGNs \citep{heck80} \--- or violent episodes of star-formation in high metallicity galaxies \citep{terl85}.

The galaxy with the next highest global N2 value is VVDS140258511
($N2_\textrm{global}=-0.509$ and $N2_\textrm{nuclear}=-0.345$). Its global and nuclear spectra show no obvious features of nuclear activity, and the N2 ratio are both $< -0.3$.

We conclude that our sample does not suffer from significant AGN contamination. The only candidate for nuclear activity is VVDS140096645. Similar conclusions were reached based on composite 1D VIMOS and SINFONI spectra of the MASSIV sample \citep[see][]{contini11}.

\subsection{Kinematics and close environment classification}\label{kin}

A kinematics and close environment classification of the MASSIV galaxies has been performed and is described in details in \cite{epinat11}. This classification is based on the $I$-band morphology, the H$\alpha$ flux maps, 
the H$\alpha$ velocity fields and their modeling assuming a rotating disk. 

In the present paper, we exploit two types of classes: a) the dynamical state of galaxies (rotating or non-rotating), 
and b) the close environment of galaxies. The first class relies i) on the agreement between the position angles of the major axis deduced from the morphology and from the kinematics, and ii) on the accuracy of the rotating disk model. 
In the case the average SNR is lower than 5, this classification is believed to be not reliable and is thus not used. The second class relies on the detection of companions in both $I$-band image and H$\alpha$ 
maps at a similar redshift than the main source. In some cases, the kinematics maps (velocity fields and velocity dispersion maps) suggest the presence of some companions about merging or along the line-of-sight. 
A confidence flag was assigned to the environment classification, ranging from A (>90\%, confident) to C ($\sim 50$\%).

We used these classifications to help us constraining the origin of the metallicity gradients in our MASSIV star-forming galaxies.

\subsection{Stellar and dark matter halo masses}

\label{sec:masses}

The stellar masses used in our study were obtained using the SED fitting technique. Stellar population synthesis models \citep{BC03} were used to match photometric and spectroscopic data from our MASSIV sample, using
the GOSSIP tool \citep{franzetti08}. A \citet{salpeter55} IMF was assumed, stellar formation timescales and the extinction parameter $E(B-V)$ were allowed to range from $0.1$ to $15$~Gyr, and from $0$ to $0.3$, respectively. 
The GOSSIP tool returned the best-fit parameters as well as a Probability Distribution Function (PDF) for each of them, following \citet{walcher08}. The median and the standard deviation of the PDF were used to recover the parameter 
estimates and their associated errors. These latter values were used for this study. The procedure is described in detail in \cite{contini11}.

The dark matter (DM) halo masses ($M_{\rm halo}$) used hereafter were computed using a spherical virialized collapse model \citep{peebles80, whiteFrenck91, mowhite02}:

\begin{equation}
\label{eq:mhalo}
M_{\rm halo} = 0.1
H_0^{-1}\mathcal{G}^{-1}\Omega_m^{-0.5}(1+z)^{1.5}V_{\rm max}^3
\end{equation}

where $\mathcal{G}$ is the universal gravitational constant, $z$ is the redshift of the galaxy and $V_{\rm max}$ is the maximum rotational velocity computed in \cite{epinat11}. In order to compute associated uncertainties, 
the Monte Carlo method used in \cite{epinat11} for the uncertainties on $V_{\rm max}$ was extended until the halo mass. The contributions of the two sources of uncertainty on the velocity (inclination and modeling) were added 
quadratically to compute the final uncertainty.

Equation \ref{eq:mhalo} makes the assumption that the plateau has been reached and that $V_{\rm max}$ traces the halo circular velocity. However, this assumption is probably not correct for non-rotating systems. 
The stellar and DM halo masses derived for our galaxies are listed in Table~\ref{metal}.

\section{Metallicity gradients}
\label{sec:zgrad}

\begin{figure}[t]
  \centering
  \includegraphics[width=\linewidth]{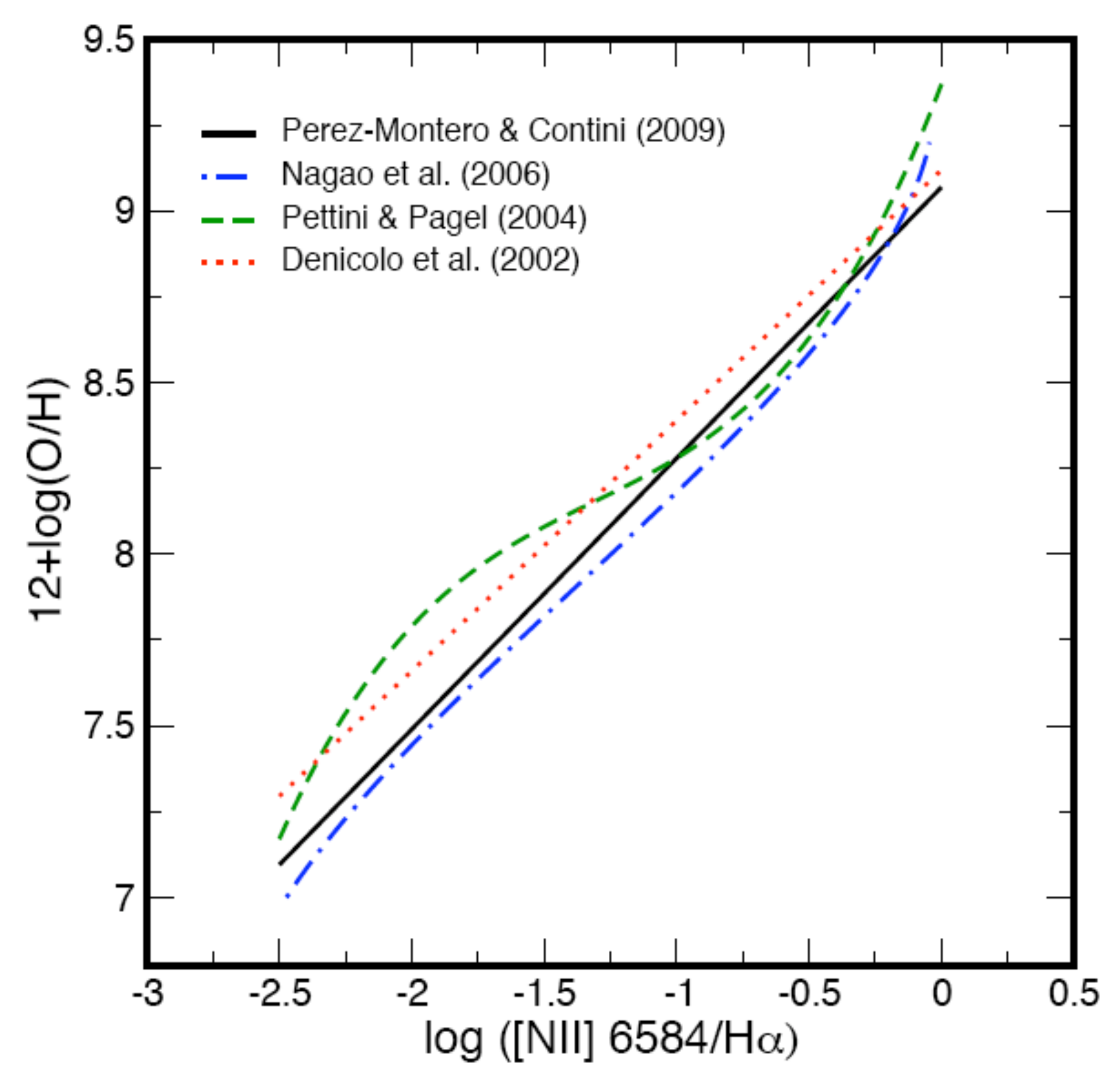}
  \caption{Comparison between different calibrations used to derive the oxygen abundance from the N2 parameter. All these calibrations 
  are consistent within their intrinsic dispersion ($\sim 0.3-0.4$ dex), especially in the range of N2 parameter ($-1.0 < {\rm N2} < -0.5$) covered by MASSIV galaxies (see Table~\ref{eml} and Figure~\ref{distrib}).}\label{compacalib}
\end{figure}

In the local Universe metallicity gradients in spiral galaxies are commonly measured \citep{pageled81, vilacostas92, considere00, pilyugin04}. These gradients are generally negative (metallicity decreasing from the center to the outer parts) and their amplitude is typically $\gtrsim-0.1$~dex kpc\un{} as traced by the metallicity in H\,\textsc{ii} regions. In the Milky Way, a gradient of $-0.07$~dex kpc\un{} is observed. 

Several physical processes can be responsible for such gradients \citep{gotz92}. Radial gas flows draining metal-rich interstellar gas from the outer parts of the galaxy into the center are believed to play a key role \citep{tinsley78}. It requires that the infall timescale of gas onto the disk is faster than the star formation
timescale. Indeed, \citet{koeppen94} has shown that the presence of both infall of pristine gas and radial flows into the disk is very efficient in creating an abundance gradient. \citet{phillipps91} further suggested that self-regulating star formation rates varying with the galactocentric distance could generate such gradients.  
In contrast, for the case of interacting galaxies, \cite{rupke10} claim that radial flows of low-metallicity gas from the merging galaxies can explain the low oxygen abundances observed at their center. In these cases, the radial mixing of gas may flatten existing metallicity gradients.

At high redshifts, metallicity gradients are harder to detect as the gas-phase metallicity of individual galaxies can only be measured with a limited accuracy. Collisional emission lines from the metals and recombination lines from hydrogen and helium are the only indicators allowing to estimate the oxygen abundance in the gas-phase, needed for a direct comparison with the local galaxies.
 
In the present study, we used the ratio of the [N\,\textsc{ii}]6584 nitrogen line to the \ha{} Balmer line as a proxy for the oxygen abundance.  Computing this particular ratio has two practical advantages: the \niia{} and \ha{} emission lines are very close in wavelength, so that {\it i)} the differential extinction due to dust attenuation can be neglected; and {\it ii)} the relative flux calibration can be considered to be  constant over the spectral range.
To derive from this ratio the gas-phase oxygen abundance, we used the calibration proposed by \citet{pmc09}:
\begin{equation}
  12 + \log(O/H) = 9.07 + 0.79\times \mathrm{N2}, \textrm{where } \mathrm{N2}=\log \frac{\mathrm{\niia{}}}
{\mathrm{H}\alpha}
\end{equation}
This calibration was computed using emission-line objects (star-forming galaxies and H\,\textsc{ii} regions) in the nearby universe with an accurate oxygen abundance obtained from the measurement of the electronic temperature. The calibration has an intrinsic scatter of 0.34 dex, mainly due to the second-order dependence of the N2 parameter on the ionisation parameter and on the nitrogen-to-oxygen abundance ratio. As shown in Figure~\ref{compacalib}, the PMC09 calibration of the N2 parameter used in this analysis is totally consistent, for both the low- and high-metallicity regime, 
with other calibrations of the same parameter found in the literature. Although the N2 parameter was calibrated by PMC09 using data with a determination of O/H abundance based only on the ``T-method", and hence not very well defined for the high-metallicity regime, the PCM09 calibration agrees very well within its dispersion limits (about $0.3-0.4$ dex) with other relations that are partially based on photoionization models or strong-line 
determinations of the metallicity \citep{denicolo02, PP04, nagaoetal06}. This is especially true in the range of N2 parameter ($-1.0 < {\rm N2} < -0.5$) covered by MASSIV galaxies (see Table~\ref{eml} and Figure~\ref{distrib}).

\subsection{Observed metallicity gradients}

\begin{figure*}[t]
  \centering
  \includegraphics[width=\linewidth]{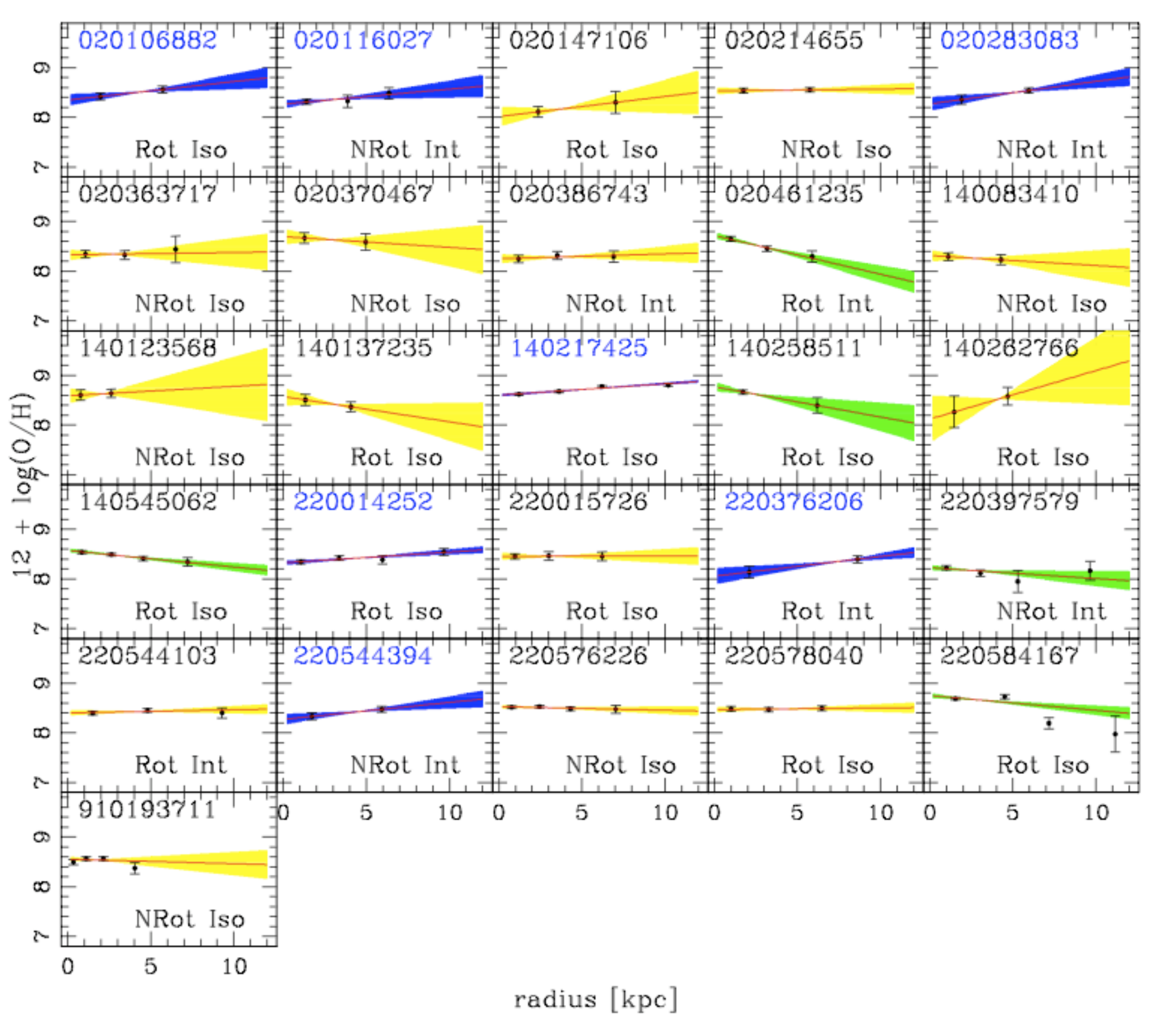}
  \caption{Metallicity gradient for the 26 MASSIV galaxies with
    spatially-resolved metallicities. The $x$-axis represents the mean
    radius (in kpc) of the region relative to the H$\alpha$
    center. The $y$-axis indicates the corresponding metallicity in
    dex ($12+\log(O/H)$). The red lines are the best fits to the data,
    taking into account the errors on the metallicities. The yellow/blue/green
    regions represent the $1\sigma$ errors associated to the
    gradients. Blue label indicates the galaxies for which the
    gradient is positive within $1\sigma$, the green ones are those
    for which it is negative within the same limits. For each galaxy
    we have indicated the dynamical (Rot=rotating disk, NRot=no rotation) 
    and environment (Iso=isolated, Int=interacting) classes.}\label{grad}
 \end{figure*}

As reported in \S\ref{el}, we measured H$\alpha$ and \niia{} line fluxes in spatially-resolved regions for 26 galaxies.  These regions, defined by \ha{} isoflux contours, are centered around the H$\alpha$ peak in each galaxy, which most often corresponds to the kinematical center of the galaxy. The widths of the annular regions were adjusted such as to have a high enough SNR for the measurement of the \niia{} line and,
consequently, the N2 ratio. This allowed to quantify the radial behaviour of the \ha{} and \niia{} lines, i.e.~of the gas-phase metallicity from the inner to the outer parts of each galaxy.  

The derived metallicity gradients and the integrated metallicities are listed in Tab~\ref{metal}. The metallicity is estimated as a
function of radius --- each region having a mean radius and a metallicity (uncertainties on metallicity estimates are dominated by measurements errors). The radius of a region corresponds to the mean radius between its outer and inner contours. The radius of a contour is approximated by the radius of a perfect circle with the same perimeter. Depending on the galaxy inclination, the intrinsic deprojected radius can be underestimated up to a factor of  $\sim 1.5$. Metallicities of the different regions in each galaxies are plotted in
\fig{grad}, along with the best fit line to the data.  The figure shows that the detected radial gradients are in general very weak, some being positives and other negatives.

\begin{table}[!ht]
\caption{Integrated metallicity ($Z = 12 + \log(O/H)$), metallicity gradient, stellar and dark matter halo masses 
of the MASSIV ``first epoch'' sample galaxies.}\label{metal}
  \centering
  \begin{tabular}{crrrr}
    \hline
    \hline
    Galaxy & $Z$ & $\Delta_r Z$ & $\log(M^\star)$ & $\log(M_{\rm h})$ \\
     & & [dex kpc\un{}] & [$M_\odot$] & [$M_\odot$] \\
     (1) & (2) & (3) & (4) & (5) \\
    \hline
    020106882 & $8.73$ & $0.037 \pm 0.026$ & $9.99_{-0.22}^{+0.22}$ & $11.59_{-0.38}^{+0.20}$\\
    020116027 & $8.42$ & $0.03 \pm 0.024$ & $10.09_{-0.23}^{+0.23}$ & $9.48_{-9.62}^{+0.38}$\\
    020126402 & $\dots$ & $\dots$ & $10.09_{-0.25}^{+0.25}$ & $\dots$\\
    020147106 & $8.49$ & $0.04 \pm 0.053$ & $10.10_{-0.13}^{+0.13}$ & $9.41_{-12.27}^{+2.87}$\\
    020149061 & $\dots$ & $\dots$ & $10.18_{-0.23}^{+0.23}$ & $11.33_{-14.19}^{+2.86}$\\
    020164388 & $8.58$ & $\dots$ & $10.13_{-0.31}^{+0.31}$ & $10.92_{-0.71}^{+0.26}$\\
    020167131 & $\dots$ & $\dots$ & $10.08_{-0.20}^{+0.19}$ & $11.58_{-0.61}^{+0.24}$\\
    020182331 & $\dots$ & $\dots$ & $10.72_{-0.11}^{+0.11}$ & $11.57_{-0.45}^{+0.22}$\\
    020193070 & $8.89$ & $\dots$ & $10.15_{-0.20}^{+0.20}$ & $11.52_{-0.42}^{+0.21}$\\
    020208482 & $\dots$ & $\dots$ & $10.17_{-0.16}^{+0.16}$ & $11.92_{-0.43}^{+0.21}$\\
    020214655 & $8.76$ & $0.0035 \pm 0.015$ & $10.02_{-0.16}^{+0.16}$ & $10.46_{-1.63}^{+0.30}$\\
    020217890 & $\dots$ & $\dots$ & $9.99_{-0.19}^{+0.19}$ & $\dots$\\
    020239133 & $\dots$ & $\dots$ & $9.89_{-0.15}^{+0.15}$ & $11.84_{-0.56}^{+0.24}$\\
    020240675 & $\dots$ & $\dots$ & $9.96_{-0.18}^{+0.18}$ & $10.26_{-13.12}^{+2.87}$\\
    020255799 & $8.80$ & $\dots$ & $9.87_{-0.16}^{+0.16}$ & $8.74_{-11.19}^{+2.46}$\\
    020261328 & $8.39$ & $\dots$ & $10.01_{-0.20}^{+0.20}$ & $11.41_{-11.45}^{+0.32}$\\
    020278667 & $\dots$ & $\dots$ & $10.28_{-0.16}^{+0.16}$ & $10.95_{-13.71}^{+2.76}$\\
    020283083 & $8.74$ & $0.045 \pm 0.026$ & $10.05_{-0.21}^{+0.21}$ & $10.56_{-0.46}^{+0.22}$\\
    020283830 & $8.61$ & $\dots$ & $10.37_{-0.17}^{+0.17}$ & $12.02_{-0.30}^{+0.18}$\\
    020294045 & $\dots$ & $\dots$ & $9.80_{-0.15}^{+0.15}$ & $12.43_{-0.55}^{+0.23}$\\
    020306817 & $\dots$ & $\dots$ & $9.76_{-0.20}^{+0.20}$ & $\dots$\\
    020363717 & $8.57$ & $0.0046 \pm 0.039$ & $9.68_{-0.20}^{+0.35}$ & $10.12_{-12.98}^{+2.86}$\\
    020370467 & $8.75$ & $-0.022 \pm 0.054$ & $10.57_{-0.14}^{+0.14}$ & $10.35_{-12.70}^{+2.35}$\\
    020386743 & $8.53$ & $0.0099 \pm 0.024$ & $9.88_{-0.20}^{+0.20}$ & $10.15_{-0.79}^{+0.26}$\\
    020461235 & $8.78$ & $-0.078 \pm 0.023$ & $10.36_{-0.15}^{+0.15}$ & $11.07_{-0.41}^{+0.21}$\\
    020461893 & $\dots$ & $\dots$ & $9.66_{-0.21}^{+0.21}$ & $10.60_{-0.54}^{+0.23}$\\
    020465775 & $\dots$ & $\dots$ & $10.12_{-0.20}^{+0.20}$ & $10.72_{-0.54}^{+0.23}$\\
    140083410 & $8.72$ & $-0.02 \pm 0.041$ & $10.07_{-0.18}^{+0.18}$ & $9.77_{-12.18}^{+2.41}$\\
    140096645 & $9.23$ & $\dots$ & $10.40_{-0.23}^{+0.24}$ & $12.75_{-15.00}^{+2.24}$\\
    140123568 & $8.75$ & $0.019 \pm 0.074$ & $9.73_{-0.39}^{+0.39}$ & $10.43_{-13.28}^{+2.85}$\\
    140137235 & $8.65$ & $-0.051 \pm 0.055$ & $10.07_{-0.29}^{+0.29}$ & $10.67_{-0.36}^{+0.19}$\\
    140217425 & $9.02$ & $0.023 \pm 0.0035$ & $10.84_{-0.17}^{+0.17}$ & $12.85_{-0.26}^{+0.16}$\\
    140258511 & $8.95$ & $-0.061 \pm 0.037$ & $10.80_{-0.48}^{+0.48}$ & $11.54_{-0.48}^{+0.22}$\\
    140262766 & $8.69$ & $0.098 \pm 0.11$ & $9.84_{-0.43}^{+0.43}$ & $11.42_{-14.27}^{+2.86}$\\
    140545062 & $8.69$ & $-0.033 \pm 0.012$ & $10.60_{-0.18}^{+0.18}$ & $12.23_{-0.58}^{+0.24}$\\
    220014252 & $8.78$ & $0.021 \pm 0.0088$ & $10.78_{-0.21}^{+0.21}$ & $11.57_{-0.50}^{+0.23}$\\
    220015726 & $8.75$ & $0.00041 \pm 0.02$ & $10.77_{-0.27}^{+0.27}$ & $12.33_{-14.76}^{+2.43}$\\
    220071601 & $\dots$ & $\dots$ & $10.81_{-0.56}^{+0.62}$ & $\dots$\\
    220148046 & $\dots$ & $\dots$ & $11.22_{-0.17}^{+0.17}$ & $10.38_{-13.23}^{+2.85}$\\
    220376206 & $8.40$ & $0.04 \pm 0.021$ & $10.67_{-0.27}^{+0.27}$ & $12.17_{-0.24}^{+0.15}$\\
    220386469 & $8.32$ & $\dots$ & $10.80_{-0.16}^{+0.16}$ & $11.30_{-0.54}^{+0.23}$\\
    220397579 & $8.38$ & $-0.022 \pm 0.02$ & $10.23_{-0.17}^{+0.17}$ & $8.18_{-9.16}^{+1.02}$\\
    220544103 & $8.89$ & $0.0063 \pm 0.012$ & $10.71_{-0.27}^{+0.27}$ & $11.62_{-0.36}^{+0.20}$\\
    220544394 & $8.55$ & $0.034 \pm 0.022$ & $10.34_{-0.23}^{+0.23}$ & $10.54_{-0.44}^{+0.21}$\\
    220576226 & $8.75$ & $-0.0073 \pm 0.01$ & $10.31_{-0.23}^{+0.23}$ & $9.77_{-10.09}^{+0.49}$\\
    220578040 & $8.64$ & $0.0031 \pm 0.013$ & $10.72_{-0.16}^{+0.17}$ & $12.48_{-14.33}^{+1.86}$\\
    220584167 & $8.53$ & $-0.03 \pm 0.013$ & $11.21_{-0.24}^{+0.24}$ & $12.31_{-0.28}^{+0.17}$\\
    220596913 & $\dots$ & $\dots$ & $10.68_{-0.30}^{+0.30}$ & $11.69_{-0.11}^{+0.08}$\\
    910193711 & $8.77$ & $-0.0091 \pm 0.028$ & $9.99_{-0.18}^{+0.42}$ & $10.56_{-0.42}^{+0.21}$\\
    910279515 & $8.60$ & $\dots$ & $10.79_{-0.14}^{+0.14}$ & $12.48_{-0.10}^{+0.08}$\\
    \hline
  \end{tabular}
  \end{table}

\begin{figure}[t]
  \centering
  \includegraphics[width=0.5\linewidth]{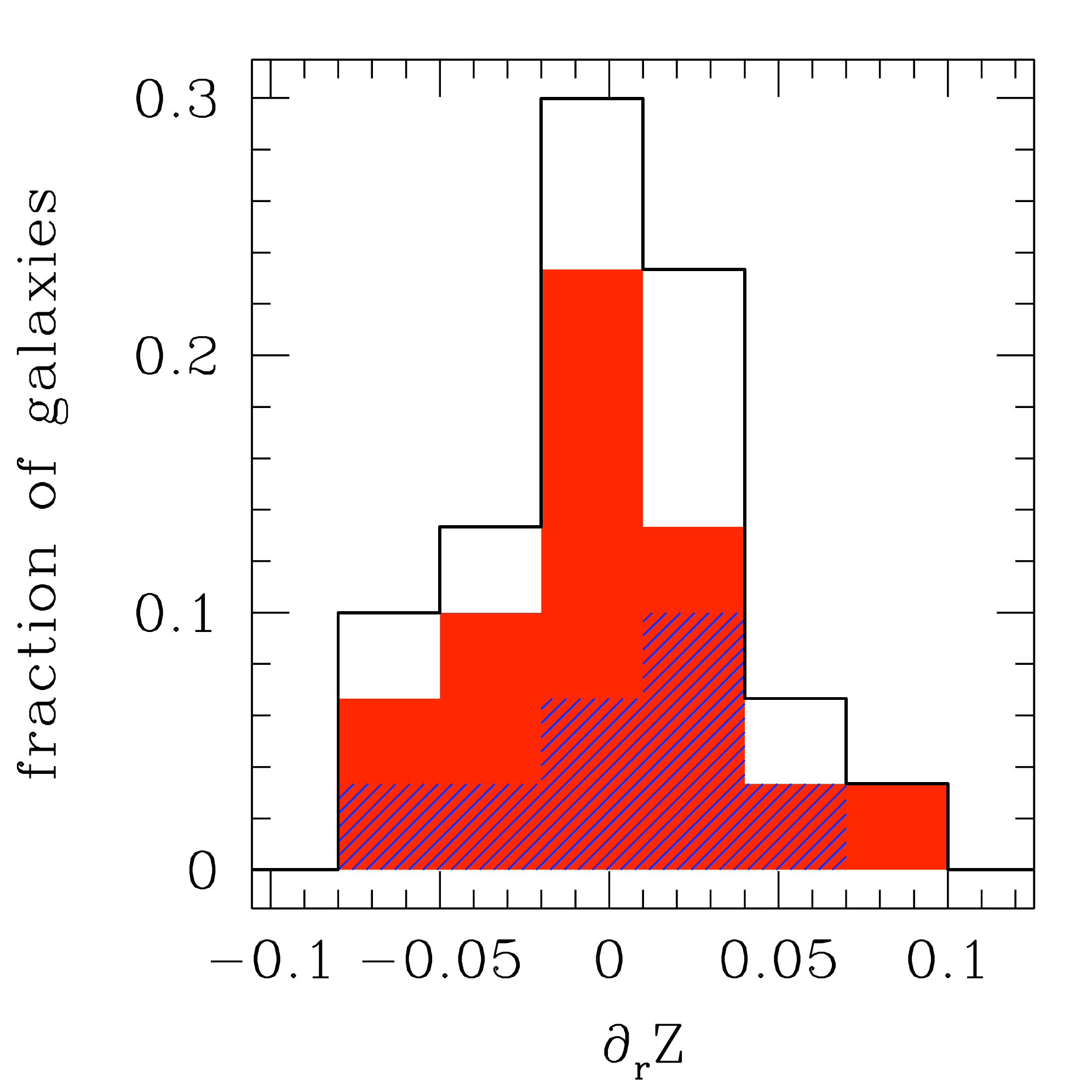}\includegraphics[width=0.5\linewidth]{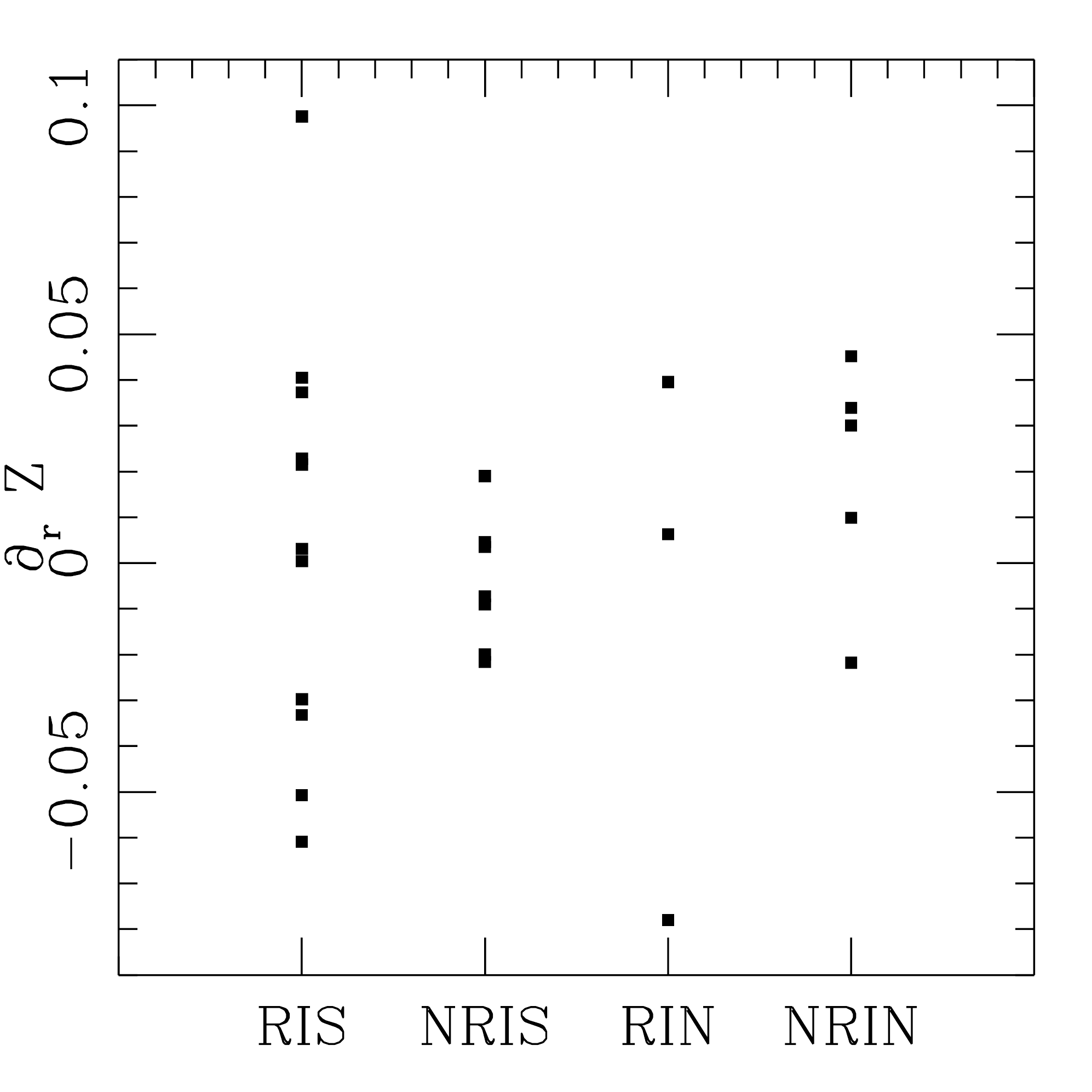}
  \caption{{\it Left}: Metallicity gradients distribution for 26
    MASSIV galaxies.  The histogram in black line represents the whole
    sample, the red and dashed blue ones respectively the isolated and
    interacting galaxies distribution. {\it Right}: Metallicity
    gradients with respect to the kinematical type (from left to
    right, Rotating Isolated, Non-Rotating Isolated, Rotating
    Interacting, Non-Rotating Interacting).}
  \label{histo_grad}
\end{figure}

\fig{histo_grad} (left panel) shows the distribution of metallicity gradients (in dex kpc\un{}) for the 26 MASSIV galaxies, for which we were able to measure \niia{} in different regions. The histograms show the full sample (in black), the isolated galaxies (in red) and the interacting ones (in dashed blue), as classified in \S\ref{kin}. For the two latter distributions, we only considered objects classified with a high or medium confidence level (A or B), excluding two isolated galaxies classified with a low confidence level C (VVDS220578040 and VVDS910193711).  In the majority of the cases, no clear gradient is detected: the median value of the total sample is $0.0040\pm 0.037$~dex kpc\un{}. The distributions for isolated (16 objects) and interacting (8 objects) galaxies show similar shapes, with median values $0.0018\pm 0.036$~dex kpc\un{} for the isolated galaxies, and $0.020 \pm 0.038$~dex kpc\un{} for interacting ones.

Interestingly though, for twelve of the 26 galaxies, a gradient is detected with a $>1\sigma$ confidence and with nearly the same proportion of positive (seven) and negative (five) gradients (see  \fig{grad}). Thus, contrary to the global trend in the local Universe, where the gas-phase metallicity of disk galaxies generally decreases with galactocentric radius, seven of our galaxies have larger metallicities in the outskirts than in the center (where the center is defined as the maximum of the \ha{} flux). 

About 1/4 of our sample galaxies thus show a positive metallicity gradient at  a $>1\sigma$ confidence level (see blue area in \fig{grad}). Of these,  VVDS220376206 displays a positive gradient with $2\sigma$ confidence, while in the galaxies VVDS140217425 and VVDS220014252 a positive gradient of $\sim0.02$ ~dex.kpc\un{} is detected with $>5\sigma$ confidence.  Among the seven galaxies, four are classified as interacting systems, while the three other are isolated. One of the isolated galaxies, the secure case VVDS140217425, appears as a ``chained-galaxy'', with large clumps in the outskirts that could be interpreted as minor mergers. About half (4/8) of the interacting galaxies have a positive metallicity gradient while it concerns only 3/16 ($\sim 20$\%) of the isolated ones. We thus tentatively conclude that the majority of the galaxies showing a positive metallicity gradient are interacting.  It is interesting to note also that among the four interacting systems, one galaxy only is classified as a rotating disk. 

On the other hand, five galaxies display clear negative metallicity gradients at a $>1\sigma$ confidence level (of which one at $>2\sigma$ and three at $>3\sigma$ confidence level; see \fig{grad} and Table \ref{metal}). Among these five galaxies, three are isolated and two are interacting. Four of these five galaxies are classified as rotating disks.

Although rare, this is not the first time that positive gradients of oxygen abundance were found.  Recently, \citet{werk10} reported a positive gradient in a local galaxy and proposed several scenarios to explain their discovery: (i) a radial redistribution of the metal-rich gas produced in the nucleus, (ii) supernovae blowing out metal-rich gas, enriching the IGM, then falling onto the outer parts of the disk, (iii) the result of a past interaction.

At high redshift, \cite{cresci10} have recently studied with SINFONI the metallicity distribution of three Lyman-break galaxies at $z\sim 3$ in the AMAZE/LSD sample. They were able to derive metallicity maps for the three galaxies with high SNR. In each case, they discovered a positive gradient, comparable to the
ones we have found. They favour the scenario in which these positive gradients would be produced by the infall of metal-poor gas into the center of the disks, diluting the gas and lowering its metallicity in the central regions. The authors claim that the discovery of positive gradients in high-redshift disks, pre-selected to be ``isolated'', is a direct evidence for cold gas accretion as a mechanism of mass assembly. Such a conclusion could be balanced arguing that a merger remnant can keep, during a transient phase with a typical timescale of $\sim 0.5$ Gyrs, an inversed metallicity gradient \citep[eg.][]{perez11, torrey11}.

In contrast, our study appears to show that among the seven detected positive gradients, only two galaxies are isolated, the others showing signs of interaction. Cold gas accretion toward the center of disks might thus not be the only process able to lower the central metallicity.

\subsection{Relations with global galaxy properties}
\label{sec:behave}

Several authors found correlations between gas-phase metallicity gradients and other global physical parameters of galaxies. The observed metallicity gradients are a function of the morphological type \citep{vilacostas92, marquez02}: they are steep in late-type spirals and almost flat for early-type spirals. Further, in the local Universe the absolute value of the gradients seems to decrease with increasing luminosity (less luminous galaxies have steeper metallicity profiles) as predicted by modeling \citep{prantzos00} and verified by observations \citep{garnett97, vanZee98}. 

Our high-redshift sample did not verify the latter correlations and we were not able to test a correlation with the morphological type, as we do not have that information for our MASSIV sample.  The right panel of \fig{histo_grad} shows no clear trend between the strength of the metallicity gradient and the kinematical type, nor any correlations with their close environment (isolated or interacting).  We can however notice that i) among the isolated objects, the non-rotating galaxies have on average flatter gradients than rotating disks, and ii) the fraction of positive gradients is higher in interacting systems compared with isolated galaxies. 

\begin{figure}[t]
  \centering
  \includegraphics[width=0.5\linewidth]{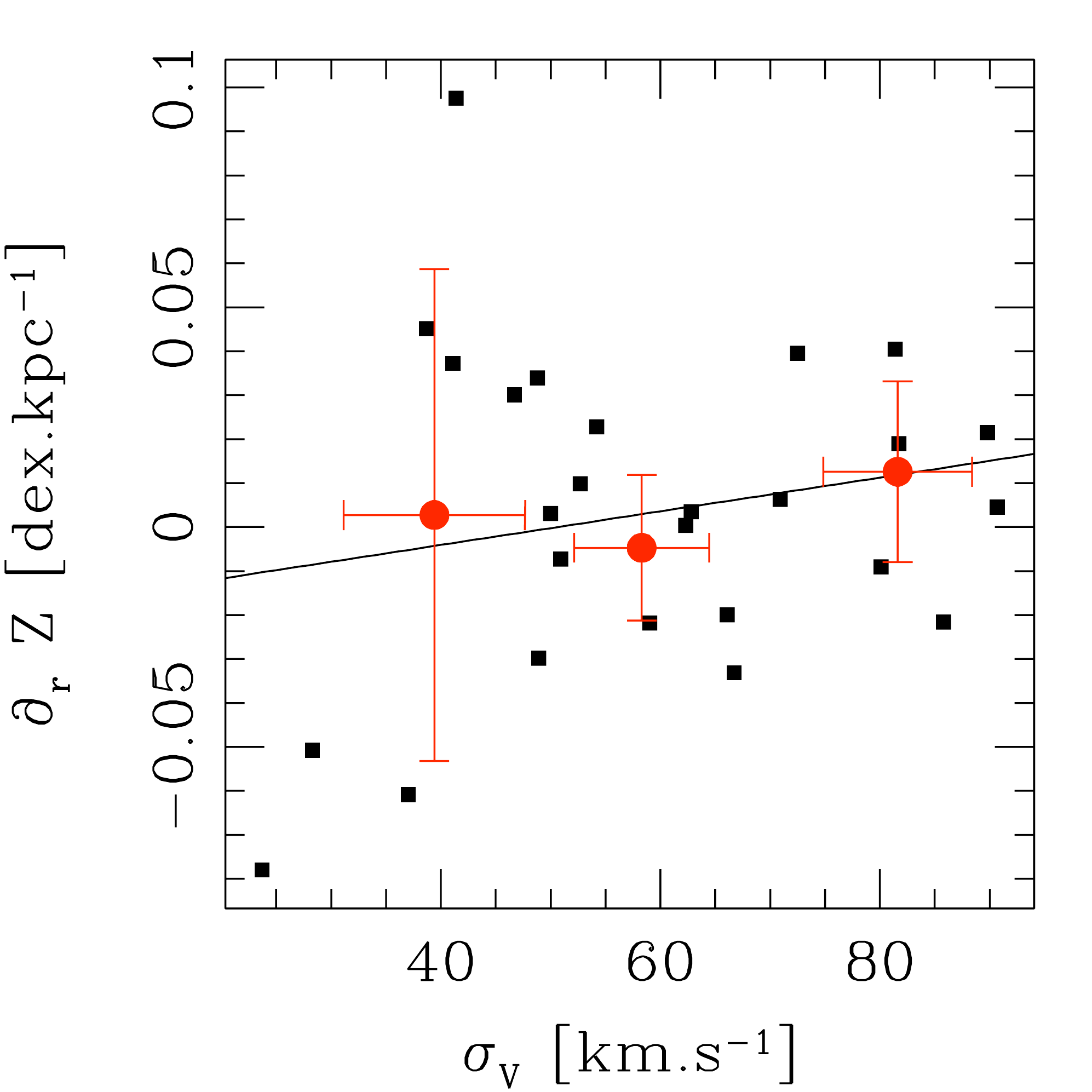}\includegraphics[width=0.5\linewidth]{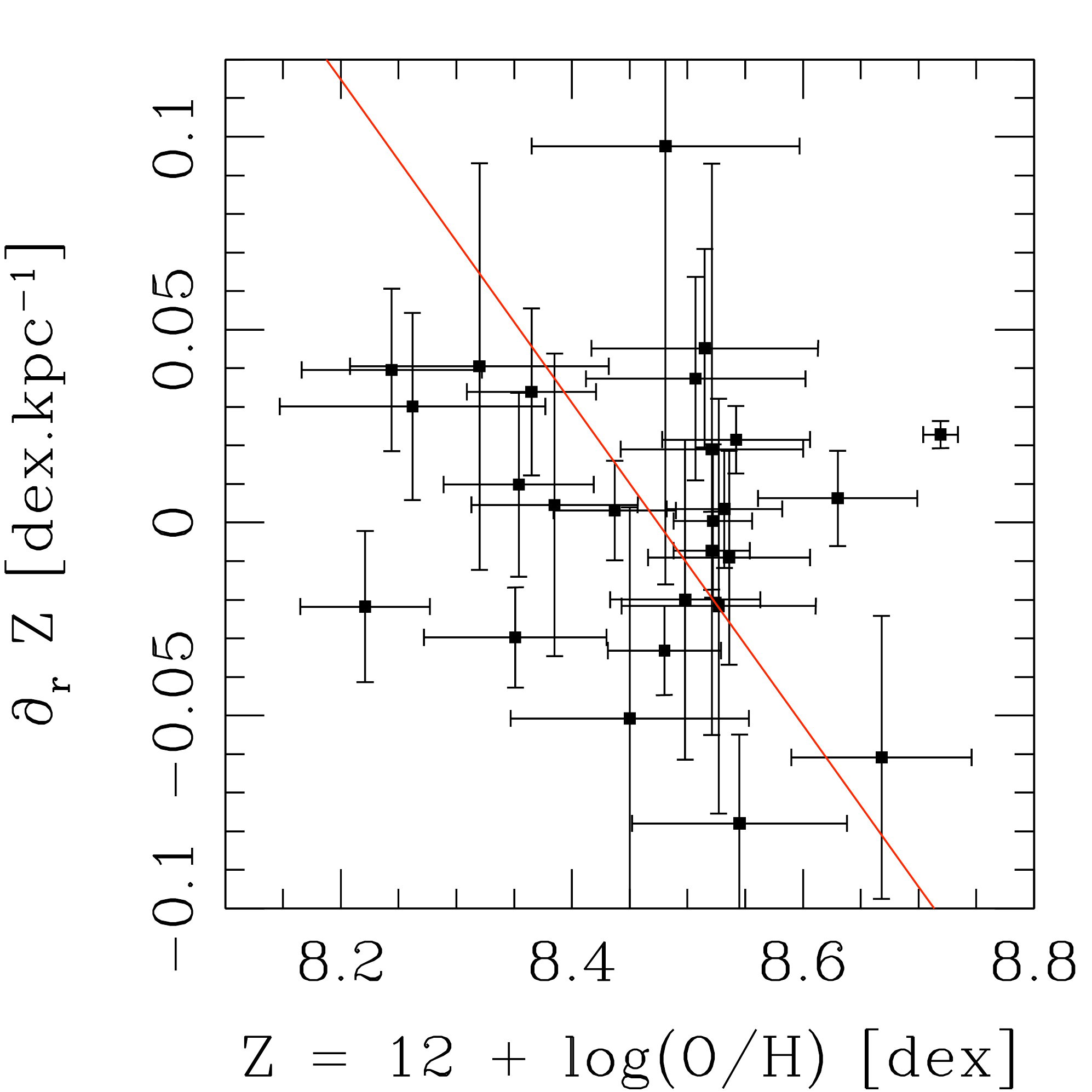}
  \caption{{\it Left}: Metallicity gradient versus the mean velocity
    dispersion of MASSIV galaxies. The black squares are the individual
    galaxies, and the red ones represent the median values for 3 bins
    of $\sigma_v$, along with the standard deviation in each bin
    represented by the error bars. {\it Right}: Metallicity gradients versus
    the integrated metallicity of each galaxy, the red line is the best
    fit to the data.}\label{fig:gradsig}
\end{figure}

One of the two weak correlations that might be present in our sample is shown in the left panel of
\fig{fig:gradsig}. The strength of the gradient seems to correlate with the velocity dispersion of the galaxy. The latter is derived on beam smearing corrected velocity dispersion maps obtained after velocity field modeling \citep[see][]{epinat11} and thus reflects the true velocity dispersion of the gas.

This apparent correlation seems to be driven by the fact that galaxies with high gas velocity dispersion show shallower, often positive, metallicity gradients. This correlation has, to our knowledge, not been observed  previously in the local universe, where the velocity dispersion in late-type objects is usually low with $\sigma_v \sim 20$~km s\un{} \citep{epinat10}. Could turbulent physical conditions in the ISM of high-redshift galaxies be at the origin of the shallow, sometimes positive gradients? This question is difficult to address here, considering the relatively low spatial resolution of our data and the scatter in \fig{fig:gradsig}. At face value, the positive gradients in our $z\sim 1.2$ galaxies might be related to the perturbed physical conditions/motions in the ISM of high-$z$ galaxies, as opposed to the continuous metallicity gradients observed in the relatively quiet ISM of the local spirals.

Finally, we observe as well a tentative anti-correlation between the metallicity gradient of each galaxy and its integrated metallicity (see \fig{fig:gradsig}, right panel).  Metallicity gradients are more frequently negative in metal-rich galaxies and more frequently positive in low-metallicity galaxies. If real, this behaviour would support the scenario in which infall of metal-poor gas from the IGM into the center of the disks drives the positive gradients. This infall of pristine gas would be able to reverse the gradient by diluting the central gas metallicity, and lowering the overall metallicity of the galaxy at the same time. The main question would remain: where does the metal-poor gas come from? Accretion of cold gas from the DM reservoir and/or interaction-triggered gas infall/capture from companions?

\subsection{Gas infall rates}

\begin{figure}[t]
  \centering
  \includegraphics[width=1\linewidth]{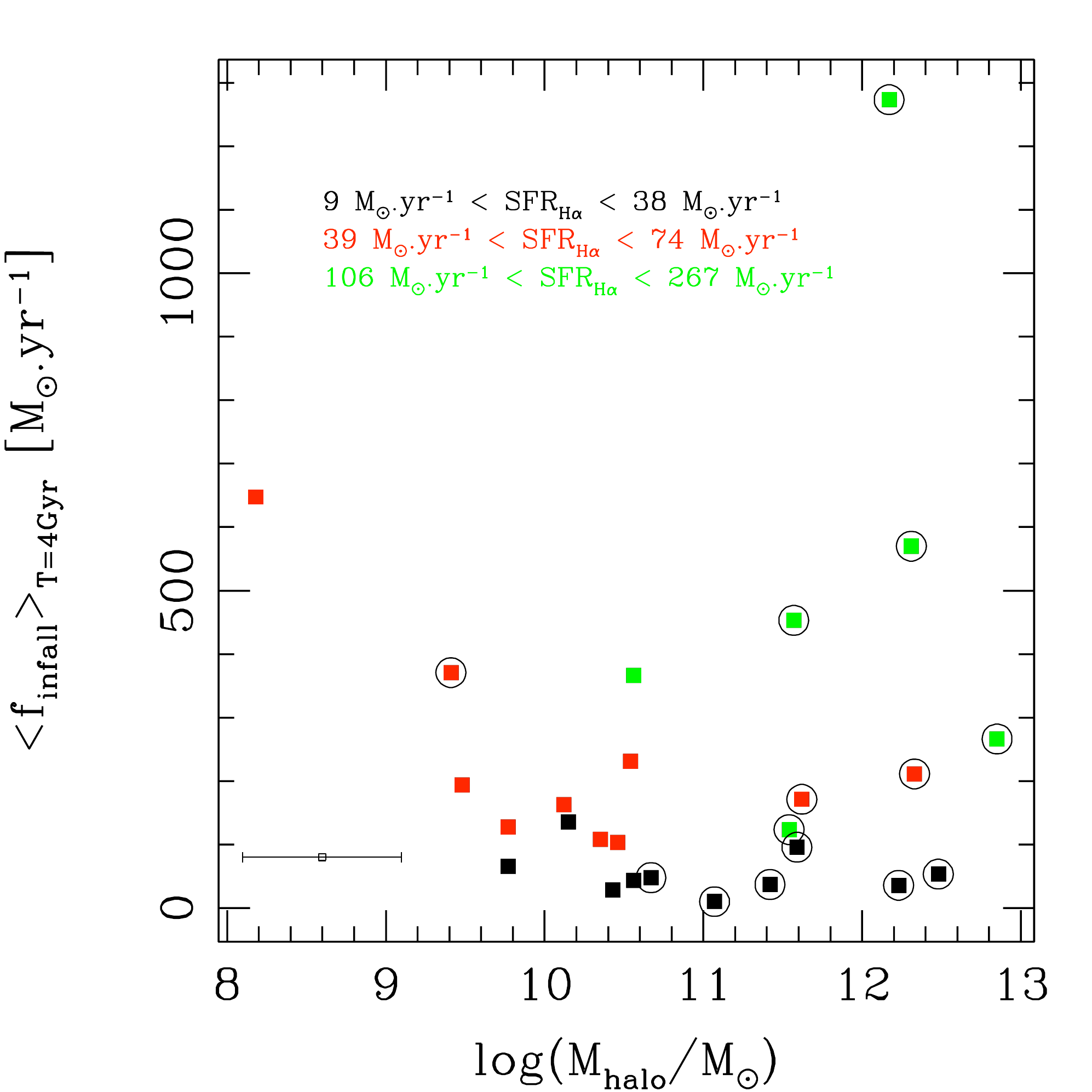}
  \caption{Mean infall rates over 4~Gyr versus the mass of the hosting
    DM halo of each MASSIV galaxies. The objects classified as
    rotators are circled in black. On the lower left corner is
    displayed a typical errorbar on the DM halo mass.}
\label{fig:infall}
\end{figure}

\begin{figure}[t]
  \centering
  \includegraphics[width=1\linewidth]{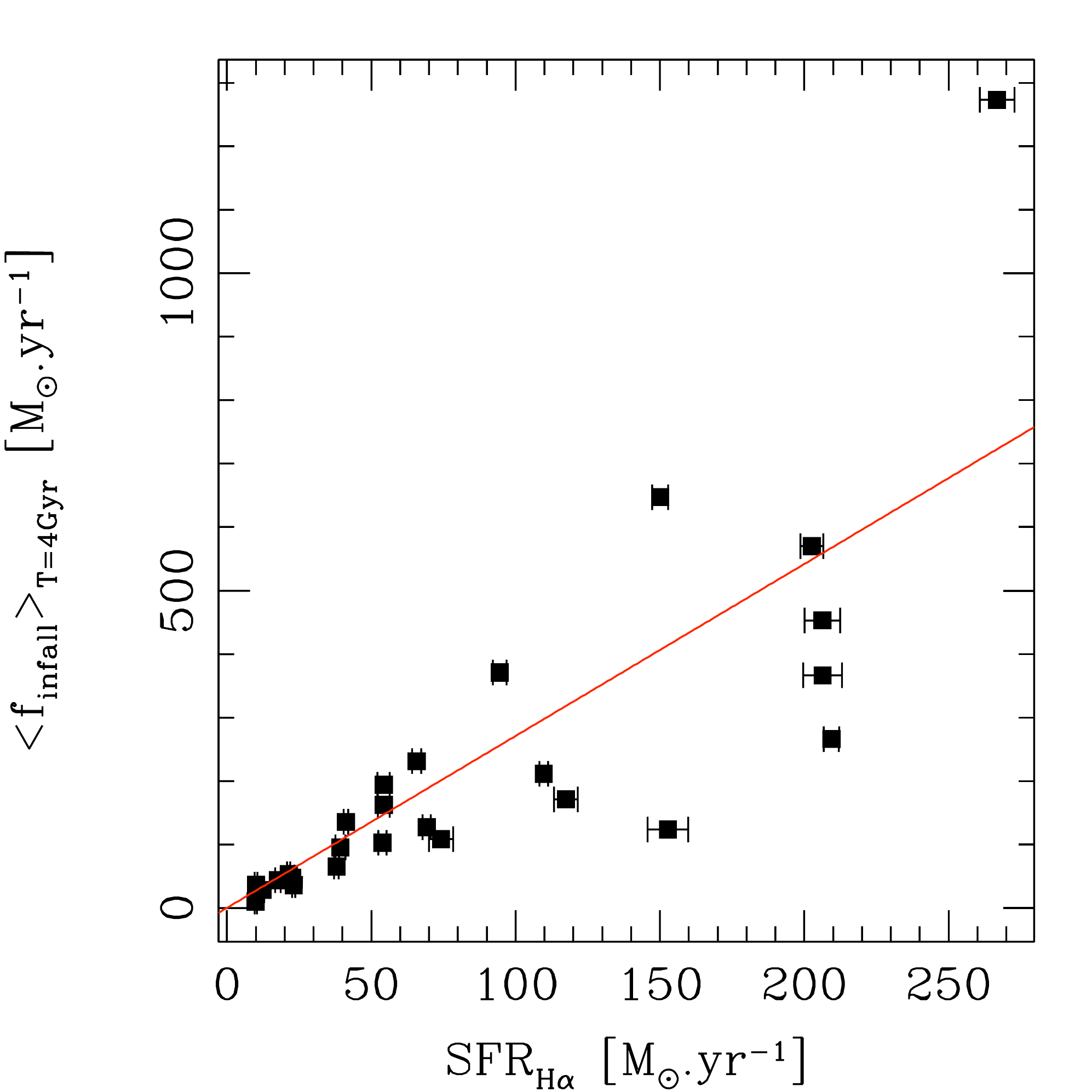}
  \caption{Mean infall rates versus the star formation rates derived
    with the \ha{} luminosity and corrected for extinction. The line is the best fit to
    the data.}
\label{fig:infall_sfr}
\end{figure}

\label{sec:toy}

As suggested already by previous studies \citep[eg.][]{werk10, rupke10}, during an interaction, metal-poor gas from the outskirts of the galaxy could 
radially flow towards the center and dilute the metallicity in the inner, high star-forming regions. At the same time, metal-enriched gas 
could be transported to the outer parts during the interaction, overall resulting in a flattening of the metallicity gradient. 

The two scenarios explaining the positive metallicity gradients \--- cold gas accretion and gas redistribution during interactions \--- have 
in common that metal-poor gas needs to be transported efficiently to the center of the objects on time scales shorter than the star formation.

We consider the possibility of infall of pristine gas onto the disk in the context of a chemical evolution model in order to explain the 
positive gradients. Our toy model for the chemical evolution of galaxies assumes (i) an instantaneous recycling approximation (IRA), (ii) the 
infall of metal-free gas onto the disks, and (iii) radial flows of gas into the disks. This leads to the following equations:

\begin{eqnarray}
\frac{\partial g}{\partial t} + \frac{\partial }{r \partial r} (r v g) & = & -(1-R)\psi(t) + f\label{toy1}\\
\frac{\partial Z}{\partial t} + \frac{\partial }{r \partial r} (r v Z ) & = & y(1-R)\psi(t) - z(1-R)\psi(t)\label{toy2}
\end{eqnarray}

where $g$ is the gas surface density, $\psi$ is the star formation rate per unit area, $R$ is the ``returned fraction'', $f$ is the infall rate 
(per unit area) of intergalactic metal-free gas, $Z = z g$ is the metal content, $y$ is the stellar yield, and $v$ is the velocity of the radial 
gas flow (positive toward the outer parts). Following a Schmidt-Kennicutt law for the star formation rate density 
($\Sigma_\psi = C g^n$, $C = 1.6\times 10^{-27}$, $n=0.7$) we solve these equations for each galaxies in our sample, assuming typical 
values for $C$, $n$, $y$, $v$ and $R$ (see above and below). After combining equations (\ref{toy1}) and (\ref{toy2}), we end up with the 
following equation allowing to derive the infall parameter $f$:

\begin{equation}
\label{eqz}
\frac{\partial z}{\partial t} + v\frac{\partial z}{\partial r} = y(1-R) C^{1/n} \psi^{1-1/n} - z C^{1/n} \psi^{-1/n} f
\end{equation}

Assuming the physical parameters involved in eq.~(\ref{eqz}) to remain constant for over $\sim 4$~Gyr (i.e.~valid for galaxies at $z \simeq 1.2$), 
we derive a mean infall rate per unit area that can be integrated since the formation epoch. To this end, we use a true yield value of $y=0.019$, 
and a returned fraction of $\sim 40$\%: $R = 0.4$ \citep{erb08}. The radial flow velocity of high-redshift galaxies is known to be higher than in local 
spirals (e.g.~$\sim 1$~km s\un{} in the case of the Milky Way). This velocity can be approximated to be the velocity towards the center of the migrating 
clumps \--- in which the star formation takes place \--- which then leads to velocities of $10-50$~km s\un{} \citep{bournaud07}. For our purpose, 
we used for each MASSIV galaxy the estimated velocity dispersion reported in \cite{epinat11}.

For the star formation rates, we used those derived from the \ha{} luminosity, corrected for extinction, and listed in Table~\ref{eml}. 
Although the \ha{} luminosity only reflects the instantaneous star formation rate, we assume it to be an average value 
since the formation epoch. As our galaxies were selected to be star-forming, this is most likely an upper limit on the average star formation.

The so-calculated global infall rates range from a few to several hundreds $\mathrm{M_\odot yr^{-1}}$. These values are high compared to the 
output of hydro-dynamical simulations of typical disk galaxies (50~\msun yr\un{}). However, the rate of infalling gas is believed to depend on the size and
mass of the host DM halo, being able to reach a few hundreds of \msun yr\un{} for the most massive galaxies \citep{keres05, erb08}. 
We have therefore derived DM halo masses for each galaxy in our sample (see Section \ref{sec:masses}) and compared it to our calculated infall 
rates (see \fig{fig:infall}). As expected there is a tendency for galaxies to accrete more gas in the most massive halos. We further divided the sample in 
three bins with respect to the (extinction-corrected) star formation rate. In that case, we noticed that galaxies with the highest SFR also have the highest 
gas-infall rate, whereas galaxies with the lowest SFR show infall rates typical for local galaxies ($\sim$50~\msun yr\un{}).

Two MASSIV galaxies deviate from the global trend in \fig{fig:infall}. VVDS220397579, located at the left end of the graph, has a very low DM 
halo mass but an important infall rate ($\sim 650~\mathrm{M_\odot yr^{-1}}$), while VVDS220376206 has by far the highest infall ($>10^3~\mathrm{M_\odot yr^{-1}}$) 
and star formation ($>250~\mathrm{M_\odot yr^{-1}}$) rates . For VVDS220397579 it is very likely that its DM halo mass has been underestimated. Indeed, this 
galaxy could be a face-on disk as the kinematical modeling returned a very low rotation velocity of $\sim 9$~km s\un{}. As the DM halo mass is directly proportional 
to the maximum rotation velocity (see equation \ref{eq:mhalo}) the DM halo mass for VVDS220397579 is most probably only a lower limit.

In \fig{fig:infall}, we have circled the objects which, according to our kinematical classification, are isolated and rotating. For those objects the DM halo 
mass is more reliable, and they appear indeed to have the largest DM halo masses. In order to investigate the relationship between the infall rate and the 
star formation rate, we show in \fig{fig:infall_sfr} the mean infall rates against the SFRs corrected for extinction. As expected from equation~\ref{eqz}, a clear linear 
correlation between these two quantities can be seen: the infall rates appear to be directly proportional to the star formation rates. The best fit to the data gives 
a slope of $2.71$ for this relation, which is defined as the infall parameter $f_i$. This was already shown to be a typical value for galaxies at these redshifts \citep{erb08}.

\section{Summary and conclusions}
\label{sec:conclu}
This paper presents the chemical abundance analysis of the ``first epoch'' sample (50 galaxies) of the MASSIV survey. Complementary analyses focused on the kinematical classification and the fundamental scaling relations can be found in two companion papers \citep{epinat11, vergani11}, as well as a description of the whole MASSIV sample \citep{contini11}.

We have been able to measure emission lines (\niia{} and \ha{}) in 34 integrated spectra of MASSIV galaxies. Within this sample, we have identified one galaxy hosting an AGN. Chemical abundance estimates could be obtained for the remaining 33 star-forming galaxies. 

For 26 galaxies, we have been able to derive a metallicity gradient, defining annular regions around the peak of the \ha{} flux (which corresponds, in most cases, to the kinematical center). While just over half of our sample is compatible with a zero metallicity gradient,  about a quarter of the galaxies shows {\it positive} gradients: the metallicity increases from the center to the outer parts of the galaxy. Among these latter (seven) galaxies, four are classified as interacting systems, one is probably a chain galaxy, and two are classified as isolated. 

Flat, more rarely positive, metallicity gradients have already been found in interacting galaxies in the local Universe. They are explained by the infall of metal-poor gas onto the central parts during the encounter \citep{werk10, rupke10}. 

Two of our galaxies showing a positive metallicity gradient are classified as isolated and do not show any sign of recent interaction. Three analogue objects were reported as isolated disk galaxies at $z\sim3$  by \citet{cresci10}. In these cases,  cold gas accretion onto the central regions of the disks seems to be the most plausible scenario. Even if tentative, there is a very interesting trend as a function of redshift. At $z\sim 3$ almost all isolated galaxies have a positive gradient, whereas this fraction drops to $\sim 15-20$\% at $\sim1.2$ and is almost equal to zero in the local universe. If cold accretion is the main process to explain the positive metallicity gradients in isolated disks, that suggests that the epoch where cold accretion dominates the mass assembly processes is at $z\geq 2$. 

We noticed in our sample a tendency for galaxies with the highest gaseous velocity dispersion to have a shallow/positive gradient, which highlights the different physical conditions observed in the ISM of high-redshift galaxies (high velocity dispersion compared to local spirals). We discovered a weak correlation between gradient and global metallicity of the galaxy: metal-poor galaxies preferentially have flat/positive gradients while metal-rich ones tend to display negative metallicity gradients. This behaviour can also be explained by the infall of
metal-poor gas onto the center of the disks, diluting the overall metallicity.

Finally, applying a simple chemical evolution model with radial flows of gas, we estimated infall rates of pristine gas onto the disks. We found values up to several hundred of \msun{} per years, and a tendency for the maximum infall rate to increases with the DM halo mass.

The analysis of the spatially-resolved metallicity of galaxies will be further extended to the full MASSIV sample, enabling a better statistics and hence a stronger interpretation in terms of galaxy assembly scenarios. The final sample will 
also allow to establish the mass-metallicity and fundamental metallicity relations at $0.9 < z < 1.8$, adding constraints on the evolutionary status of star-forming galaxies at high redshifts.

\begin{acknowledgements}
  We wish to thank L\'eo Michel-Dansac and Fr\'ed\'eric Bournaud for
  their help and useful comments on this work. We thank also the referee for useful suggestions. This work has been partially supported by the
CNRS-INSU and its Programme National Cosmologie-Galaxies (France) and by the french ANR grant ANR-07-JCJC-0009. DV acknowledges the support from the INAF contract PRIN-2008/1.06.11.02.

\end{acknowledgements}

\newpage \bibliographystyle{aa} \bibliography{gradz}

\end{document}